\documentclass[12pt]{article}

\usepackage[english]{babel}

\usepackage[letterpaper,top=2cm,bottom=2cm,left=3cm,right=3cm,marginparwidth=1.75cm]{geometry}
\usepackage{adjustbox}
\usepackage{amsmath}
\usepackage{graphicx}
\usepackage{threeparttable}
\usepackage{siunitx}
\usepackage[hidelinks]{hyperref} %, allcolors=blue [colorlinks=true]
\usepackage[utf8]{inputenc} % 允许使用UTF-8编码
\usepackage[T1]{fontenc}    % 使用8位T1编码字体
\usepackage{amsmath}        % AMS数学公式支持
\usepackage{amssymb}        % 额外的数学符号
\usepackage{amsfonts}       % AMS字体
\usepackage{graphicx}       % 插入图片
\usepackage{booktabs}       % 专业质量的表格
\usepackage{multirow}       % 合并表格中的多行
\usepackage{enumitem}       % 自定义列表格式
\usepackage{url}            % 简单的URL排版
\usepackage{ragged2e}       % 文本自动换行和对齐
\usepackage{tabularx}       % 自适应列宽的表格
\usepackage{float}          % 改进浮动对象的处理
\usepackage{threeparttable} % 含注释的表格
\usepackage{longtable}      % 跨页的长表格
\usepackage{caption}        % 自定义标题
\usepackage{subcaption}     % 子图
\usepackage{rotating}       % 旋转对象
\usepackage{appendix}       % 附录支持
\usepackage{siunitx}        % 更好的数值和科学记数法
\usepackage{listings}
\usepackage{xcolor}
\usepackage{rotating}
\usepackage{bm}
\usepackage[round,authoryear]{natbib}
\usepackage{setspace}  % Added package for spacing
\usepackage{authblk}  % author/affiliation commands
\hypersetup{
  colorlinks=true,      % 启用彩色超链接
  linkcolor=blue,       % 目录和内部链接为蓝色
  urlcolor=blue,        % URL 链接为蓝色
  filecolor=blue,       % 文件链接为蓝色
  citecolor=blue,       % 文献引用链接为蓝色
}
\usepackage{pdflscape}
\title{What influenced the lack of diversity in CSR after the company's losses: evidence from topic modeling\footnote{Ruiying Liu, Yuchi Li and Zhanli Li contribute equally}}
\author[1]{Ruiying Liu\thanks{Liu: No.182, Nanhu Avenue, East Lake High-tech Development Zone, Wuhan, Hubei Province, P.R. China, 430073. E-mail: liuruiying@stu.zuel.edu.cn}}
\author[2]{Yuchi Li\thanks{Li: No.45, Zengguang Road, Haidian District, Beijing, P.R. China, 100048, E-mail: 2290411015@stu.culr.due.cn}}
\author[1]{Zhanli Li\thanks{Li: No.182, Nanhu Avenue, East Lake High-tech Development Zone, Wuhan, Hubei Province, P.R. China, 430073, E-mail: lizhanli@stu.zuel.edu.cn}}
\date{}
\affil[1]{Wenlan School of Business, Zhongnan University of Economics and Law}
\affil[2]{School of Labor Economics, China University of Labor Relations}

\begin{document}

\maketitle
\vspace{0.1cm}
\begin{abstract}
The diversity of corporate social responsibility (CSR) disclosure is a crucial dimension of corporate transparency, reflecting the breadth and resilience of a firm’s social responsibility. Using CSR reports of Chinese A-share firms from 2006 to 2023, this paper applies Latent Dirichlet Allocation (LDA) to extract topics and quantifies disclosure diversity using the Gini-Simpson index and Shannon entropy. Regression results show that corporate losses significantly compress CSR topic diversity, consistent with the “slack resources hypothesis.” Both external and internal governance mechanisms mitigate this effect: higher media attention, stronger executive compensation incentives, and greater supervisory board shareholding attenuate the loss–diversity penalty. Results are robust to instrumental variables estimation, propensity score matching, and placebo tests. Heterogeneity analyses indicate weaker effects in firms with third-party assurance, those disclosing work safety content, large firms, and those in less competitive industries. Our study highlights the structural impact of financial distress on non-financial disclosure and provides practical implications for optimizing CSR communication, refining evaluation frameworks for rating agencies, and designing diversified disclosure standards.
\end{abstract}
\vspace{0.2cm}


\newpage

\section{Introduction}
The idea of corporate social responsibility (CSR) emerged in the early twentieth century from a re-examination of firm–society relations within industrial capitalism, centering on how to balance profit and the public good within a single ethical frame. In 1953, ~\citet{bowen2013social} offered the first systematic treatment by defining CSR as a managerial orientation that guides policies, decision processes, and courses of action toward social welfare. Subsequently, ~\citet{carroll1979three} proposed the “pyramid” model that stratifies responsibility into economic, legal, ethical, and philanthropic layers, institutionalizing obligations “beyond profit.” In the 1980s, ~\citet{freeman2010strategic} broadened CSR’s boundary via the stakeholder perspective, requiring firms to consider employees, consumers, suppliers, communities, and the natural environment in addition to shareholders. Amid globalization, controversies over labor exploitation and environmental damage intensified this paradigm shift. Ignoring stakeholder interests not only impairs reputation and invites market boycotts but also pushes CSR from a “cost center” toward a strategic lever for long-term performance ~\citep{donaldson1995stakeholder}. By 2000, the United Nations Global Compact had nudged CSR practice toward quasi-regulation by emphasizing international principles on human rights, labor, environment, and anti-corruption to mitigate globalization’s negative externalities and support the architecture of sustainable development governance ~\citep{rasche2009toward}.

In emerging economies such as China, government-led disclosure has gradually become a key governance tool under the dual pressures of ongoing economic integration and tightening ecological constraints. Following the Shenzhen Stock Exchange’s Guidelines for the Social Responsibility of Listed Companies in 2006, some firms were brought under mandatory or quasi-mandatory CSR reporting obligations. This laid a regulatory foundation for subsequent institutionalization ~\citep{chen2018effect}. This policy move, consonant with the goal of building a “Harmonious Society,” sought to alleviate distributional tensions and environmental bottlenecks through corporate responsibility practice. Early reports, however, were highly dispersed due to immature standards and uneven supervisory approaches. Content often remained at the level of philanthropy or compliance narratives and fell short of capital markets’ growing demand for decision-useful nonfinancial information ~\citep{li2010corporate}. In parallel, research and market evidence document rising investor preference for nonfinancial dimensions such as CSR and ESG because of their incremental information for risk identification and long-horizon valuation ~\citep{cohen2015nonfinancial,amel2018and}. Yet in local practice, firms continue to navigate competitive pressure, institutional gaps, and regulatory heterogeneity. CSR texts at times function as instruments of political survival rather than a strategic nucleus, hampering the shift from symbolic narration to embedded governance ~\citep{li2010corporate}.

Heightened salience of climate and sustainability has further shifted disclosure from “compliance response” toward “proactive communication,” with firms cultivating stakeholder expectations by reducing information asymmetry ~\citep{du2010maximizing}. In carbon governance, information on emissions, energy efficiency, and green investment commands increasing attention. After China announced the goals of carbon peaking and carbon neutrality in 2021, climate-related content has been embedded more systematically and with stronger strategic attributes ~\citep{lu2024regulating}. As a communication vehicle, the CSR report not only records social performance but also advances systematization and transparency in risk and opportunity management ~\citep{moravcikova2015csr}. Since the 2008 global financial crisis, the triple bottom line (TBL) has been widely adopted to balance economic, social, and environmental dimensions ~\citep{velte2022meta}. To improve comparability, GRI standards have become dominant in practice. ~\citet{widiarto2009social} propose social disclosure rating schemes to gauge report quality and completeness, while the UN-driven ESG architecture and the ISSB’s reporting standards further calibrate reporting through quantifiable metrics and an investor lens ~\citep{kazmierczak2022literature}.

The rise of media scrutiny has profoundly shaped both the content and cadence of CSR communication. High-profile exposés on environmental pollution and labor rights directly affect public and investor assessments. For example, after a widely reported 2010 incident concerning supply-chain labor practices at a firm in southern China, the company intensified disclosure to repair reputation ~\citep{zhang2020media}. Although completeness, relevance, and reliability have improved ~\citep{stuart2023defining}, reports still tend at times toward surface-level presentation and fail to capture local specificities ~\citep{koh2023corporate}. Topics such as rural revitalization and the Belt and Road Initiative reflect unique cultural and policy contexts but remain underrepresented in existing frameworks ~\citep{jamali2018corporate}. Meanwhile, the shadows of legitimacy management and impression management persist. More pages do not necessarily translate into measurability of performance, and investors continue to face difficulty in discerning real outcomes ~\citep{cho2015csr}.

The rapid ascent of ESG discourse has injected new momentum into transparency but has also bred concerns about homogeneity. Compared with the broader CSR paradigm, ESG centers more on measurable E–S–G indicators and investor orientation. Firms increasingly treat annual or standalone reports as channels to court ESG-linked capital. In practice, they sometimes substitute “ESG texts” for more commitment-laden CSR narratives, steering information architectures toward convergence under quantification. Among domestic rating agencies, for instance, Huazheng (CSI) emphasizes environmental factors (e.g., carbon intensity), whereas Wind assigns greater weight to governance. Differing indicator scopes and weightings alter rating sensitivities and can induce firms to align with particular evaluation systems, diluting place-based sustainability efforts and making disclosure look like a compliance-check response rather than substantive communication focused on long-term value ~\citep{tsang2024rise,christensen2022corporate,berg2022aggregate}. On the regulatory side, the lack of arrangements that encourage diversified presentation and narrative innovation—combined with assessment-driven pressures—further amplifies “compliance-is-enough” incentives and slows progress toward comprehensive and truthful disclosure ~\citep{lu2024regulating}.

Traditional textual analyses of CSR reports have often relied on bespoke dictionaries. In a setting without unified, stringent disclosure norms, researcher-constructed lexicons are vulnerable to selection bias. Subjective curation rarely exhausts the range of CSR topics, constraining findings by design. Against this backdrop, “topic diversity” is not only a core dimension of the breadth of CSR practice but also a key indicator of disclosure transparency. Especially relative to the full-market corpus, the topic diversity within a given firm’s CSR report helps neutralize selective-disclosure noise and enhances comparability and credibility. Given substantial heterogeneity in firms’ reporting conventions—often “speaking different dialects”—we apply latent Dirichlet allocation (LDA) to CSR texts for A-share listed companies during 2006–2023 to model latent topic structures. Relative to TF–IDF-type methods, LDA excels at capturing contextual associations and topic heterogeneity ~\citep{goloshchapova2019corporate,szekely2017can}. At the measurement layer, we compute Gini coefficients and Shannon entropy from document-level topic distributions to depict concentration and evenness, thereby evaluating disclosure diversity. The post-2006 institutionalization of CSR in China’s A-share market supplies a continuous and ample textual sample and reflects the co-evolution of “mandated requirements” and “localized responses” ~\citep{chen2018effect}.

Within this measurement framework, we return to the relation between financial outcomes and responsibility communication. Much prior work emphasizes positive effects—for instance, CSR can reduce default risk ~\citep{boubaker2020does}—yet systematic evidence on how losses reshape the topic structure of disclosure remains scarce. As a canonical form of financial distress, losses can operate through two channels: (i) resource constraints prompt cuts to non-core topics, reducing topic diversity ~\citep{harymawan2021financially}; (ii) to repair reputation and stabilize expectations, management may broaden information to strengthen responsiveness ~\citep{zhang2021can}. Evidence from developing economies further suggests that CSR spending can exacerbate financial pressure under overinvestment or when external opinions diverge ~\citep{farooq2021impact,tarighi2022corporate}. In China’s context, it is intuitive—and not uncommon—that firms treat CSR during loss periods more as a crisis-management tool than as a strategic hub.

This study’s potential contributions are fourfold and interrelated. Methodologically, by quantifying topic structure with LDA and mapping it to diversity via Gini and entropy, we offer a dynamic, data-driven alternative to dictionary approaches and pivot measurement from intensity to structure. Theoretically, we integrate legitimacy logic with resource constraints into a unified framework to explain how financial distress suppresses or reshapes information architectures. Empirically, drawing on a large A-share sample, we estimate the effect of losses on topic diversity and examine moderating roles of external media attention and internal governance—executive pay and supervisory board shareholding—alongside heterogeneity analyses to strengthen interpretability. In practice, we provide evidence to inform firms’ optimization of responsibility communication, rating agencies’ refinement of evaluation scopes, and regulators’ design of incentives for diversified disclosure standards, thereby helping stakeholders address governance challenges in the local context and enhance long-term sustainability performance.

\section{Hypothesis Development}
\subsection{Corporate Losses and CSR Diversity}

As economic organizations, firms prioritize the stability and safety of operations. According to the slack resources hypothesis proposed by ~\citet{preston1997corporate}, discretionary (non-core) CSR activities should be undertaken only after continuity in core business is secured; when financial conditions are sound and slack resources exist, firms may allocate funds to philanthropy, green innovation, and other non-core activities ~\citep{waddock1997corporate, ullmann1985data}. This view aligns with resource-constraint theory: under tight resources, rigid expenses (e.g., raw materials, wages) take precedence, while spending on non-core items such as corporate social responsibility  is more adjustable ~\citep{orlitzky2003corporate}. When firms face losses or external financing constraints, slack resources shrink and deferrable outlays like CSR are cut back ~\citep{surroca2010corporate}. Hence, from a resource-supply perspective, the intensity of CSR investment closely follows financial condition; only financially healthy firms retain the discretion to fund non-core activities.

Legitimacy theory, however, holds that firms must also sustain social legitimacy by complying with social norms, moral standards, and stakeholder expectations ~\citep{suchman1995managing}. Legitimacy both underpins access to resources and motivates CSR engagement and disclosure. To obtain social support, mitigate external conflict, and stabilize capital-market expectations, firms typically respond with transparent CSR disclosure ~\citep{long2025does}. In financial distress, a tension emerges: resource scarcity pushes cuts to non-core spending, whereas legitimacy pressure calls for maintaining sufficient disclosure and responsibility practices.

Under this tension, firms often choose a “minimum-cost compliance” strategy. They rely on standardized report formats to cover core, verifiable, material topics to meet compliance and avoid the reputational and regulatory risks of silence, while reducing topic diversity to conserve resources. By contrast, blindly expanding topics under resource scarcity can induce non-substantive disclosure, easily read as symbolic management or “greenwashing,” and may deepen legitimacy crises ~\citep{kuzey2023financial}. Thus, firms in distress tend to keep necessary compliance disclosure but reduce topic diversity.

We acknowledge potential countervailing mechanisms and boundary conditions. Some firms may use high-profile CSR disclosure as a countercyclical signal to soften negative interpretations of distress. In tightly regulated or reputation-sensitive industries, firms may be compelled to maintain higher topic diversity. Such cases usually rely on stronger external oversight or internal governance. Overall, the interplay between resource constraints and legitimacy pressure leads firms in distress toward “minimum-cost compliance,” compressing topic diversity.

\medskip
\noindent H1: Relative to profitable firms, loss-making firms are less able to sustain higher CSR topic diversity; their CSR reports exhibit lower topic diversity.

\subsection{Moderating Mechanisms}

Combining the slack resources hypothesis and legitimacy theory, loss-making firms tend to cut deferrable spending and reduce topic diversity. External oversight and internal governance can, however, alter managerial choices and affect resource allocation and disclosure. We examine three mechanisms that may mitigate the “minimum-cost compliance” tendency under losses: media attention (external oversight), executive compensation (internal incentives), and supervisory board shareholding (internal monitoring).

\subsubsection{External Oversight by the Media}

As a key external governance force, the media filters and amplifies information, placing corporate actions under public scrutiny and increasing visibility and transparency ~\citep{dyck2008corporate,bushee2010role}. Greater media attention raises the minimum disclosure standard required to sustain social legitimacy; disclosures once deemed sufficient may now appear inadequate ~\citep{suchman1995managing}. Media monitoring also increases the likelihood that symbolic or selective disclosure will be identified and amplified. When loss-making, cutting topics due to resource pressure can be labeled as impression management or “greenwashing” ~\citep{kim2015greenwash}. Negative coverage transmits quickly through public opinion and investor sentiment, heightening financing constraints and valuation discounts ~\citep{tetlock2007giving}. Consequently, higher media attention increases the marginal cost of compressing CSR topic diversity in loss-making firms and pushes them to maintain greater, more balanced, and verifiable topic diversity to preserve legitimacy and reduce financing risk.

\medskip
\noindent H2a: Under losses, higher media attention is associated with higher CSR topic diversity.

\subsubsection{Internal Incentive Effect of Executive Compensation}

Executive pay, a core element of internal governance, shapes managers’ resource allocation and disclosure strategies through incentive intensity and career concerns ~\citep{jensen1990performance}. Higher pay often comes with stricter performance pressure and accountability risk; if disclosure missteps trigger market backlash, highly paid managers face larger expected losses and thus favor completeness and verifiability. Higher pay also tends to coincide with larger and more complex organizations, where managers face more diverse stakeholders and stricter external standards. In such settings, CSR disclosure is treated as a necessary cost to sustain legitimacy and relational capital, not a dispensable slack item ~\citep{gabaix2008has}. Prior work further shows that in firms with stronger governance, executive compensation and environmental or social performance are more likely to be positively related, reflecting attention to nonfinancial performance ~\citep{surroca2010corporate}. We therefore expect stronger pay incentives to reduce loss-making firms’ tendency to compress CSR topic diversity.

\medskip
\noindent H2b: Under losses, higher executive compensation is associated with higher CSR topic diversity.

\subsubsection{Internal Monitoring Effect of the Supervisory Board}

The supervisory board, a central internal monitoring institution, oversees managerial decisions, ensures compliance in financial reporting, and protects shareholder interests, thereby mitigating agency conflicts and improving transparency ~\citep{denis2003international}. In China’s two-tier board structure, supervisory boards monitor both the board of directors and executives, including financials and disclosure ~\citep{liu2017corporate}. When supervisory board members hold more shares, their wealth is more tightly linked to long-term firm value. This raises the expected benefits of monitoring and narrows tolerance for short-termism. Shareholding heightens sensitivity to reputational and compliance risks and increases monitoring effort, making “bare-minimum compliance with narrowed topic diversity” harder to pass internal review. For loss-making firms inclined to cut CSR spending, higher supervisory board shareholding raises the cost of compressing topic diversity and encourages broader, more balanced disclosure.

\medskip
\noindent H2c: Under losses, higher supervisory board shareholding is associated with higher CSR topic diversity.

\medskip
In sum, the tug-of-war between resource supply and legitimacy pressure drives loss-making firms toward “minimum-cost compliance,” reducing CSR topic diversity. Media attention, executive compensation, and supervisory board shareholding—via external oversight, internal incentives, and internal monitoring—raise the marginal cost of narrowing topic diversity. These main and moderating effects jointly form the study’s theoretical framework and testable hypotheses.

\section{Data}
\subsection{Data Sources and Sample Selection}
We use Chinese A‐share listed companies from 2006–2023 as our research sample. Financial data come from the China Stock Market \& Accounting Research (CSMAR) database and the China Research Data Services Platform (CNRDS). CSR report texts are scraped from CNINFO  using Python scripts. We set 2006 as the starting year because it marks a key inflection point in the institutionalization of CSR in China: that year CSR was first written into the Company Law as a statutory obligation.\footnote{The \emph{Company Law of the People’s Republic of China} was amended on October 27, 2005 and took effect on January 1, 2006. Article 5 states: “In its business activities, a company shall comply with laws and administrative regulations, observe social ethics and business ethics, be honest and trustworthy, accept supervision by the government and the public, and assume social responsibility.” This provision significantly increased the number of firms issuing CSR reports starting in 2006.} This regulatory change led to a rapid rise in the number of CSR reports, providing ample cross-sectional and time-series variation to examine the relationship between corporate losses and CSR topic diversity.

To preserve data integrity while focusing on the core relationship, we only drop observations with missing CSR topic diversity measures or key financial variables. We retain *ST and ST firms as well as financial-industry observations: the former are more likely to be in financial distress and face stricter regulation—the focal context of this study—and their CSR disclosure is distinctive; the latter (financial firms) interact closely with other sectors and their CSR reports contain sector-specific topics that should not be ignored. This minimal filtering strategy maintains data quality while avoiding selection bias from excessive deletion. The final baseline sample includes 2,257 unique firms and 13,797 firm–year observations.

\subsection{Dependent variable: CSR Topic Diversity}
We first collect listed firms’ CSR reports and, based on the Jieba segmentation library, expand the user dictionary with entries from MBA Zhiku and each firm’s full name and common abbreviations. We then standardize case, remove punctuation, and drop common stop words to obtain stable token representations.

Latent Dirichlet Allocation (LDA) is a generative probabilistic model used to discover latent semantic topics in a corpus \citep{blei2003latent}. Its basic assumptions are as follows. Let the corpus contain $D$ documents, and let the word sequence of document $d$ be $\bm{w}d=(w{d1},\dots,w_{dN_d})$. For a given number of topics $K$:
\begin{align}
  \bm{\theta}_d &\sim \mathrm{Dirichlet}(\bm{\alpha}), \qquad d=1,\dots,D, \label{eq:doc-topic}\\
  z_{dn}\mid \bm{\theta}_d &\sim \mathrm{Categorical}(\bm{\theta}_d), \qquad n=1,\dots,N_d,\\
  \bm{\phi}_k &\sim \mathrm{Dirichlet}(\bm{\beta}), \qquad k=1,\dots,K, \label{eq:topic-word}\\
  w_{dn}\mid z_{dn}=k,\,\bm{\phi}_k &\sim \mathrm{Categorical}(\bm{\phi}_k).
\end{align}
Here, $\bm{\theta}d$ is the topic distribution of document $d$, $z{dn}\in{1,\dots,K}$ is the latent topic assignment for word $w{dn}$, and $\bm{\phi}_k$ is the word distribution of topic $k$. The priors $\bm{\alpha}$ and $\bm{\beta}$ control sparsity and smoothness, respectively.

For model selection and hyperparameter tuning, we use validation log-perplexity as the outer objective. Denote the hyperparameter vector as
\begin{equation}
  \bm{x}=\big(K,\;\texttt{passes},\;\alpha,\;\beta\;\big)
\end{equation}
After fitting on the training set, we compute on an independent validation set $\mathcal{D}{\mathrm{val}}$:
\begin{equation}
  y(\bm{x}) = \mathrm{log\_perplexity}(\mathcal{D}_{\mathrm{val}};\bm{x})
  = -\frac{1}{N_{\mathrm{val}}}\sum_{w_i\in\mathcal{D}_{\mathrm{val}}}\log p(w_i\mid \bm{x}),
\end{equation}
and solve
\begin{equation}
  \min_{\bm{x}\in\mathcal{X}} y(\bm{x}). \label{eq:outer-opt}
\end{equation}
Perplexity is defined as
\begin{equation}
  \mathrm{Perplexity}(\mathcal{D}_{\mathrm{val}}) = \exp\!\left( -\frac{1}{N_{\mathrm{val}}}\sum_{w_i\in\mathcal{D}_{\mathrm{val}}}\log p(w_i\mid \bm{x}) \right),
\end{equation}
so minimizing $\mathrm{log_perplexity}$ is equivalent to minimizing $\mathrm{Perplexity}$.

To efficiently explore the hyperparameter space, we adopt Optuna’s default sampler—the Tree-structured Parzen Estimator (TPE) \citep{bergstra2011algorithms,bergstra2013making}. Given historical observations $\mathcal{D}={(\bm{x}i,y_i)}{i=1}^n$, TPE chooses a threshold $y^\star$ (typically the $\gamma$-quantile of $y$, $\gamma\in(0,1)$) and splits observations into “good” and “bad” sets:
\begin{align}
  \mathcal{G} &= \{(\bm{x}_i,y_i): y_i \le y^\star\},\qquad
  \mathcal{B} = \{(\bm{x}_i,y_i): y_i > y^\star\}.
\end{align}
It then estimates the conditional densities
\begin{equation}
  \ell(\bm{x}) := p(\bm{x}\mid y \le y^\star), \qquad
  g(\bm{x}) := p(\bm{x}\mid y > y^\star)
\end{equation}
nonparametrically: continuous/integer dimensions use kernel density estimation with a Gaussian kernel in this study. Rather than modeling $p(y\mid\bm{x})$ directly, TPE maximizes the expected improvement
\begin{equation}
  \mathrm{EI}(\bm{x})=\int \max\{0,\,y^\star-y\}\,p(y\mid \bm{x})\,dy
\end{equation}
which can be rewritten to maximizing the ratio
\begin{equation}
  \rho(\bm{x})=\frac{\ell(\bm{x})}{g(\bm{x})}. \label{eq:tpe-ratio}
\end{equation}
In practice, TPE samples candidates from $\ell(\bm{x})$ and selects the point with the largest $\rho(\bm{x})$, balancing exploitation of high-probability good regions and exploration of uncovered areas. Because our objective is to minimize $y(\bm{x})$ (log-perplexity), $y^\star$ and the set split treat smaller losses as “good,” consistent with \eqref{eq:outer-opt}.

Based on the tuned model, we estimate for each CSR report its topic distribution vector
\begin{equation}
{\theta}_d = (p_{d1},p_{d2},\dots,p_{dK}), 
\end{equation}
where $p_{di}$ denotes the probability weight of report $d$ on topic $i$, and $\sum_{i=1}^K p_{di} = 1$. Because the corpus covers all disclosed CSR report texts, firms necessarily allocate finite space and attention across the $K$ candidate topics when reporting. Assuming truthful disclosure without deliberate fabrication, firms typically emphasize some issues more than others. Such variation in coverage across topics provides a feasible basis to measure CSR \emph{topic diversity}.

From the perspective of probability distributions, concentration on a few topics implies lower diversity; a more even spread implies higher diversity. We quantify this feature using two complementary measures:
\begin{itemize}
    \item \textbf{Gini-Simpson index}:
In industrial organization, market concentration is commonly summarized by the Herfindahl–Hirschman Index (HHI), the sum of squared market shares. Interpreting topic proportions $p_{di}$ as “shares,” the complement to that concentration measure provides a natural diversity metric—the Gini-Simpson index \citep{simpson1949measurement}.
    \begin{equation}
        Gini = 1 - \sum_{i=1}^K p_{di}^2,
    \end{equation}
where a larger value indicates a more even distribution and thus higher topic diversity.

    \item \textbf{Shannon entropy}:
From information theory \citep{shannon1948mathematical}, it captures uncertainty/information content:
    \begin{equation}
        Entropy = - \sum_{i=1}^K p_{di} \log p_{di},
    \end{equation}
where a larger value indicates a more uniform distribution and hence stronger topic diversity.
\end{itemize}

Taken together, these two indicators describe CSR report topic diversity from the angles of “inequality” and “uncertainty,” providing complementary measures for the subsequent empirical analysis.

\begin{figure}[H]
  \centering
  \begin{minipage}{0.48\textwidth}
    \centering
    \includegraphics[width=\linewidth]{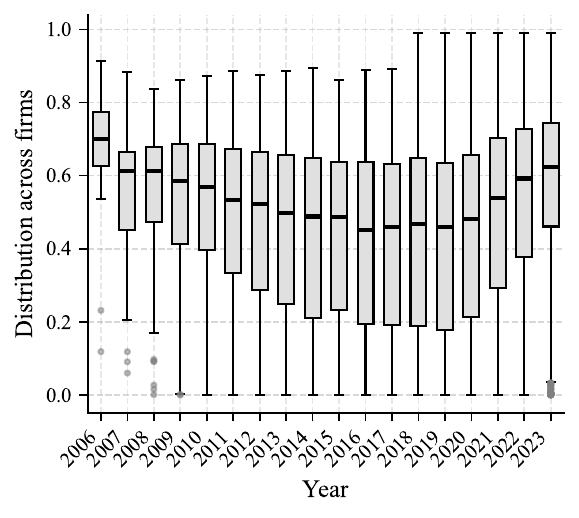}
    \caption{Gini-Simpson}
    \label{fig:gini}
  \end{minipage}
  \hfill
  \begin{minipage}{0.48\textwidth}
    \centering
    \includegraphics[width=\linewidth]{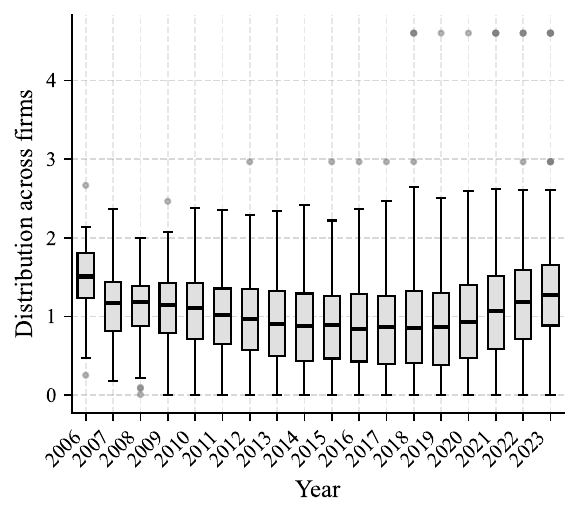}
    \caption{Shannon Entropy}
    \label{fig:entropy}
  \end{minipage}
  \caption*{\footnotesize\emph{Notes: }These two figures show annual box plots of firm-level CSR topic diversity—Gini–Simpson on the left and Shannon entropy on the right. For each year, the box spans the interquartile range from the first to the third quartile, and the horizontal line inside marks the median. The whiskers extend to the most extreme values within 1.5 times the interquartile range from the quartiles. Dots beyond the whiskers indicate outlying observations.}
 
\end{figure}

For ease of interpretation of the CSR topic diversity measures derived from the LDA topic distributions, we present annual box plots for the Gini–Simpson and Shannon entropy metrics, as illustrated in Figure~\ref{fig:gini} and Figure~\ref{fig:entropy}. The box plots display, for each year, the median, interquartile range, and outliers, providing a visual summary of year-to-year variation. The two metrics exhibit similar shapes at the annual level, indicating that they describe the same notion of “evenness of topic coverage.” Accordingly, we report both measures in parallel and use them as mutual robustness checks in the analyses that follow.

\subsection{Methodology Construction}
Building on the research design above, we specify the following baseline regression in which CSR topic diversity is lagged by one period. This choice reflects two considerations: (i) a firm's financial performance (e.g., profit/loss) in year $t$ is more likely to influence its CSR report disclosed in year $t\!+\!1$, this operation is also used by~\citep{li2025esg}; (ii) lagging the CSR diversity variable helps mitigate endogeneity concerns arising from potential reverse causality.
\begin{equation}
\label{baseline}
    CSR\_Div_{i,t+1} = \beta_0 + \beta_1 Loss_{i,t} + \beta_2 Controls_{i,t} +\lambda_i+\gamma_t+ \mu_{it}
\end{equation}
where \(CSR\_Div_{i,t+1}\) denotes the CSR topic diversity of firm \(i\) in year \(t+1\), measured using LDA topic modeling. We employ two alternative proxies for CSR diversity: \(CSR\_Div\_gini_{i,t+1}\), based on the Gini coefficient of topic distributions, and \(CSR\_Div\_ent_{i,t+1}\), based on the entropy of topic distributions. \(Loss_{i,t}\) is a dummy equal to 1 if firm \(i\) reports a loss in year \(t\) and 0 otherwise. \(Controls_{i,t}\) is a vector including firm size (Size), leverage ratio (Lev), operating cash flow (Cashflow), fixed asset ratio (Fixed), the largest shareholder’s ownership (Top1), board size (Board), and firm value (TobinQ). Table~\ref{tab:variable_description} provides detailed variable definitions. The model further includes firm fixed effects \(\lambda_i\) and year fixed effects \(\gamma_t\); \(\mu_{it}\) is the error term.

\begin{center}
\textbf{[Insert Table~\ref{tab:variable_description} Here]}
\end{center}

\subsection{Descriptive Statistics}
Table~\ref{tab:descriptive} lists the descriptive statistics of the main variables. The mean and standard deviation of CSR\_Div\_gini are 0.4768 and 0.2523, respectively, indicating that the average topic diversity of CSR reports is moderate. The mean of CSR\_Div\_ent is 1.0089, with a standard deviation of 0.5913. Notably, the 75th percentiles of the dependent variables are 0.6823 and 1.4381, respectively, which suggests that most companies adopt a conservative disclosure strategy. For the key independent variable Loss, the mean and standard deviation are 0.1074 and 0.3097, indicating that approximately 10.74\% of the observations are loss-making. Meanwhile, the moderating variables Mt, Salary, and Sshrrat have few missing values.

\begin{center}
\textbf{[Insert Table~\ref{tab:descriptive} Here]}
\end{center}

\section{Empirical results}

\subsection{Baseline Regression}
To evaluate the effect of corporate losses on the topic diversity of CSR reports, we estimate the model in Equation~\eqref{baseline} with firm and year fixed effects and cluster standard errors at the firm level. This specification controls for unobserved, time-invariant firm heterogeneity and common shocks while addressing potential within-firm correlation and heteroskedasticity, thereby improving the reliability of the estimates. Table~\ref{tab:CSR_FinancialDistress} reports the baseline results: columns (1)–(2) use the Gini-based measure and columns (3)–(4) use the entropy-based measure. Regardless of whether controls are included, the coefficient on Loss is significantly negative at the 1\% level, indicating a robust association between losses and lower CSR topic diversity. This finding confirms H1: relative to profitable firms, loss-making firms exhibit significantly lower CSR topic diversity.

\begin{center}
\textbf{[Insert Table~\ref{tab:CSR_FinancialDistress} Here]}
\end{center}

\subsection{The Moderating Effect of Media Attention}
Consistent with the resource–legitimacy framework, media attention shapes external oversight and thereby influences disclosure choices by loss-making firms. When firms incur losses, resource constraints raise the marginal cost of expanding topics, pushing them toward “minimum-cost compliance”; higher media visibility, however, increases the expected reputational costs of symbolic or selective disclosure, strengthening legitimacy pressure and inducing “remedial communication”~\citep{dyck2008corporate,zyglidopoulos2012does}. In particular, negative coverage tends to associate insufficient disclosure with “greenwashing,” amplifying reputational and financing risks~\citep{miller2006press}.

Empirically, we use CNRDS counts of firm-level news by tone (positive, neutral, negative) and construct interaction terms (Loss $\times$ Mt) to test whether media attention moderates the loss–diversity relationship. Table~\ref{tab:news_moderation} reports the results: columns (1)–(2) correspond to positive media attention, columns (3)–(4) to neutral attention, and columns (5)–(6) to negative attention, using CSR\_Div\_gini and CSR\_Div\_ent, respectively, as outcome variables. The interaction coefficients are generally significantly positive, implying that media attention mitigates the adverse effect of losses on CSR topic diversity. The moderating effect is strongest for negative media attention, followed by neutral, and weakest for positive attention. This pattern reflects the more potent disciplinary role of negative coverage: when firms report losses, adverse news draws the scrutiny of stakeholders and compels firms to broaden disclosure to prevent further diffusion of negative public opinion~\citep{dyck2008corporate,zyglidopoulos2012does}. By comparison, positive coverage entails weaker oversight and correspondingly smaller effects on topic diversity~\citep{miller2006press}. These results support H2a.

\begin{center}
\textbf{[Insert Table~\ref{tab:news_moderation} Here]}
\end{center}

\subsection{The Moderating Effect of Executive Compensation}
In resource-constrained settings, whether managers expand nonfinancial disclosure depends critically on incentives and career concerns. Performance-sensitive pay ties managerial payoffs to external evaluation, heightening the perceived costs of disclosure missteps~\citep{jensen1990performance,conyon2011executive}. More complete and balanced disclosure improves the information environment, eases financing constraints, and strengthens relational capital, thereby supporting long-term firm value~\citep{healy2001information}. Accordingly, higher compensation—especially for core decision-makers—should increase managerial preference for verifiable and well-covered disclosure, reducing the likelihood of being labeled as “greenwashing,” while organizational complexity and stakeholder pressure also tend to co-move with higher pay~\citep{gabaix2008has}.

Using CSMAR executive pay data, we adopt two measures—top-three executive pay (Salarytop3) and total management pay (Salarysum)—and interact each with Loss to test moderation. Table~\ref{tab:pay_moderation} shows that, with few exceptions, the interaction terms are significantly positive, with the Salarytop3 interaction displaying more consistent robustness across both diversity measures. These results indicate that stronger pay incentives—particularly for the core management team—mitigate the negative effect of losses on CSR topic diversity by encouraging broader and more balanced disclosure to manage legitimacy and financing risks, supporting H2b.

\begin{center}
\textbf{[Insert Table~\ref{tab:pay_moderation} Here]}
\end{center}

\subsection{The Moderating Effect of Board Shareholding}
Supervisory board shareholding is an important internal monitoring mechanism during financial distress. In China’s two-tier governance structure, the supervisory board independently oversees executives and the board of directors; stronger ownership ties heighten monitoring incentives and align interests with long-term value~\citep{denis2003international,liu2017corporate}. Based on CSMAR data on supervisory board shareholding (Sshrrat), we construct the interaction term (Loss $\times$ Sshrrat) to test moderation.

Table~\ref{tab:supervisory_share_moderation} shows that the interaction coefficients are positive and significant at the 5\% and 1\% levels, indicating that higher supervisory board shareholding attenuates the negative impact of losses on CSR topic diversity. Greater ownership strengthens oversight of managerial behavior and promotes broader, more balanced CSR disclosure~\citep{denis2003international}, thereby validating H2c.

\begin{center}
\textbf{[Insert Table~\ref{tab:supervisory_share_moderation} Here]}
\end{center}

\subsection{Robustness Check}
\subsubsection{Winsorization}
To guard against results being driven by extreme values, we winsorize all continuous variables at the 1\% level on both tails. As shown in columns (1)-(2) of Table~\ref{tab:robustness1.1}, whether or not we winsorize, the coefficient on \emph{Loss} remains significantly negative at the 1\% level. This indicates that outliers do not materially affect our findings and supports the reliability of the results.
Although our baseline already includes key controls, omitted external factors may still affect the content of CSR reports. In particular, whether a firm follows the GRI (Global Reporting Initiative) standards can shape the normativeness and breadth of CSR disclosure. To address potential heterogeneity from differing reporting standards, we add an indicator for GRI adherence as a control. Columns (1) and (2) of Table~\ref{tab:robustness1} show that, after including GRI, the coefficient on \emph{Loss} remains significantly negative at the 1\% level, indicating that corporate losses continue to reduce CSR topic diversity. The effect thus survives this additional control.

\subsubsection{Exclusion of Special Observations}
Firms designated ST or *ST are subject to exchange risk warnings and differentiated oversight; their disclosure incentives and information environments differ from the regular sample. Financial firms also exhibit systematic differences in capital structure, regulatory regime, and CSR reporting norms. To ensure the generality of our conclusions, we exclude these special observations. As reported in columns (3)-(4) of Table~\ref{tab:robustness1.1}, the coefficient on Loss remains significantly negative at the 1\% level, with a magnitude close to the baseline estimates, suggesting that these observations have limited influence on the results.

\begin{center}
\textbf{[Insert Table~\ref{tab:robustness1.1} Here]}
\end{center}

\subsubsection{Addition of Control Variables}
Although our baseline already includes key controls, omitted external factors may still affect the content of CSR reports. In particular, whether a firm follows the GRI (Global Reporting Initiative) standards can shape the normativeness and breadth of CSR disclosure. To address potential heterogeneity from differing reporting standards, we add an indicator for GRI adherence as a control. Columns (1)-(2) of Table~\ref{tab:robustness1} show that, after including GRI, the coefficient on Loss remains significantly negative at the 1\% level, indicating that corporate losses continue to reduce CSR topic diversity. The effect thus survives this additional control.

\subsubsection{Alternative Independent Variable (Roa)}
Because Loss is a binary variable, it cannot reflect the intensity of profitability or distress and may be highly correlated with other financial variables, raising multicollinearity concerns for the baseline. We therefore re-estimate the model using return on assets (Roa) in place of Loss to assess how financial condition relates to CSR topic diversity. ROA is defined as net profit divided by average total assets; as a continuous measure, it captures profitability more precisely and is mathematically negatively related to Loss, providing a useful validation of our main conclusion. Columns (3)-(4) of Table~\ref{tab:robustness1} show that the baseline finding continues to hold. Overall, the ROA coefficient is significantly positive at the 10\% and 5\% levels, implying that stronger profitability is associated with higher CSR topic diversity—consistent with our main result that loss-making firms reduce CSR topic diversity.

\begin{center}
\textbf{[Insert Table~\ref{tab:robustness1} Here]}
\end{center}

\subsubsection{Instrumental variable estimations}

Although we use next-period CSR diversity and a two-way fixed-effects specification—which together mitigate several endogeneity concerns—the persistence of corporate policies means that unobserved factors $Z$ may still drive CSR diversity over time, yielding $Cov(Loss,\mu)\neq 0$. We therefore adopt an instrumental-variables (IV) approach and estimate via two-stage least squares (2SLS), treating Loss as endogenous. A valid instrument must be correlated with Loss but affect CSR topic diversity only through Loss (exclusion restriction).

We use the interaction between $t{+}2$ CSR diversity measures and contemporaneous profitability as external instruments. Specifically, we construct three IVs:
\begin{equation}
    IV_1 = CSR\_Div\_gini_{t+2} \times Roa
\end{equation}
\begin{equation}
    IV_2 = CSR\_Div\_ent_{t+2} \times Roa
\end{equation}
\begin{equation}
IV_3 = (CSR\_Div\_gini_{t+2} + CSR\_Div\_ent_{t+2}) \times Roa
\end{equation}

where $CSR\_Div\_gini_{t+2}$ and $CSR\_Div\_ent_{t+2}$ are the $t{+}2$ CSR diversity measures, and $Roa$ is the $t$-period return on assets.

\emph{Relevance.} As a direct measure of profitability, Roa is mechanically and strongly negatively associated with Loss, a relationship frequently exploited in empirical work; for example, ~\citet{acharya2017measuring} use profitability-type indicators to forecast financial risk and losses. Interacting Roa with $t{+}2$ CSR diversity captures forward-looking variation that strengthens the predictive content for Loss, akin to ~\citet{campello2010real}, who interact future investment opportunities with current financial indicators as IVs and document high first-stage $F$ statistics.

\emph{Exogeneity.} The $t{+}2$ CSR diversity measures are future values and, conditional on controls and fixed effects, should not directly affect the $t$-period CSR topic diversity; rather, they reflect longer-run structural forces (e.g., industry norms or policy trajectories) rather than contemporaneous shocks at $t$, and they do not create a feedback channel into the current outcome. Related identification strategies appear in ~\citet{gormley2014common}, who use $t{+}1/t{+}2$ industry shocks as IVs, arguing that future events can satisfy the exclusion restriction when they are external to the current dependent variable. Similarly, ~\citet{bennouri2018female} employ interactions with future ESG-style measures, and ~\citet{jo2012causal} use future CSR constructs as instruments, both emphasizing that $t{+}2$ values do not directly shift current performance.

We evaluate instrument strength and identification using standard diagnostics. As reported in Table~\ref{tab:2sls_results}, the Kleibergen–Paap rk LM statistic rejects the null of underidentification at the 1\% level. The Kleibergen–Paap rk Wald $F$ statistics (210.099, 172.512, and 187.061) far exceed the Stock–Yogo critical value of 16.38 for a 10\% maximal IV size, ruling out weak identification concerns. Columns (1), (4), and (7) present the first-stage regressions, where the instruments load on \emph{Loss} with the expected negative and statistically significant coefficients. The second-stage estimates show that \emph{Loss} remains significantly negative at the 1\% level, indicating that even after addressing endogeneity via IV, corporate losses are associated with lower CSR topic diversity. These results reinforce the baseline evidence.

\begin{center}
\textbf{[Insert Table~\ref{tab:2sls_results} Here]}
\end{center}
\subsubsection{Propensity score matching method}
Even though we include rich firm-level and external controls, CSR topic diversity may relate to covariates in a nonlinear fashion. Such functional form misspecification (FFM) can bias $\hat{\beta_1}$. Following ~\citet{shipman2017propensity}, we implement propensity score matching (PSM) to reduce dependence on functional form and alleviate endogeneity due to FFM by reweighting toward comparable observations.

We proceed as follows. First, we split firms into a treated group (loss-making) and a control group (non-loss) and estimate propensity scores via a logit using all baseline controls as covariates. Second, within the average treatment effect framework, we perform 1:1 nearest-neighbor matching with a caliper of 0.05 and restrict matches to the common support. Post-matching, the average treatment effects on the treated (ATT) equal \,-0.0229 and \,-0.0414, both statistically significant. Balance diagnostics indicate that standardized \%bias across covariates falls below 5, and $t$-tests fail to reject equality between treated and matched controls, supporting covariate balance and the maintained overlap. Table~\ref{tab:robustness_psm} reports the matched-sample estimates, which continue to show a significantly negative association between Loss and CSR topic diversity, consistent with our main results.

\begin{center}
\textbf{[Insert Table~\ref{tab:robustness_psm} Here]}
\end{center}

\subsubsection{Placebo tests}
A remaining concern is that the baseline relationship between \emph{Loss} and CSR topic diversity might be a placebo driven by unobserved factors rather than a causal channel. Following ~\citet{dhaliwal2011voluntary} and ~\citet{chetty2009salience}, we conduct placebo tests by randomly reassigning Loss across firm identifiers within each year to create a pseudo treatment, and we re-estimate the baseline model on each placebo sample. We repeat this procedure 1{,}000 times.

\begin{figure}[H]
    \centering
    \includegraphics[width=0.7\linewidth]{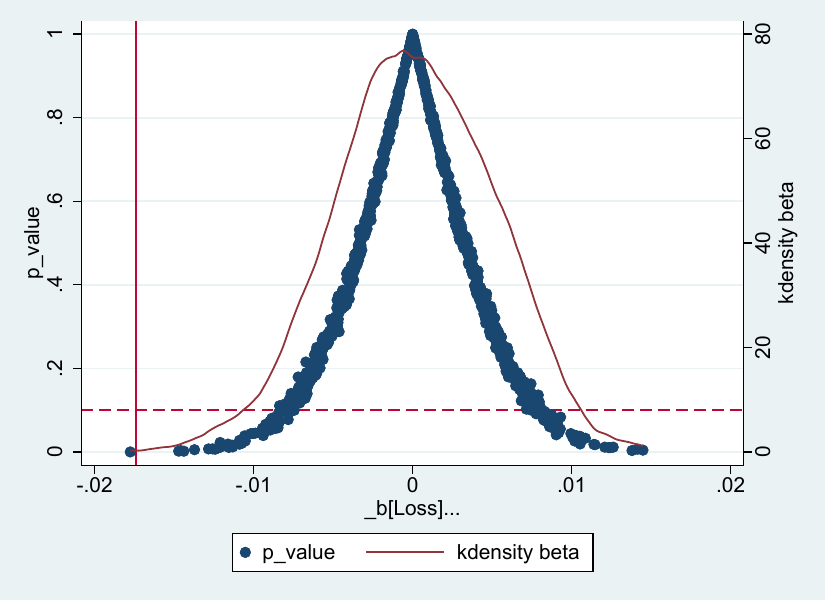}
\caption{Placebo test (CSR\_Div\_gini).}
\caption*{\footnotesize\emph{Note:} The x-axis plots coefficients from 1,000 placebo regressions with randomized Loss; dark blue dots are the corresponding $p$-values (left axis), and the red curve is the kernel density (right axis). The vertical solid line marks the baseline estimate $\hat{\beta}=-0.0174$, and the horizontal dashed line indicates $p=0.10$.}
    \label{fig:placebo_gini}
\end{figure}

\begin{figure}[H]
    \centering
    \includegraphics[width=0.7\linewidth]{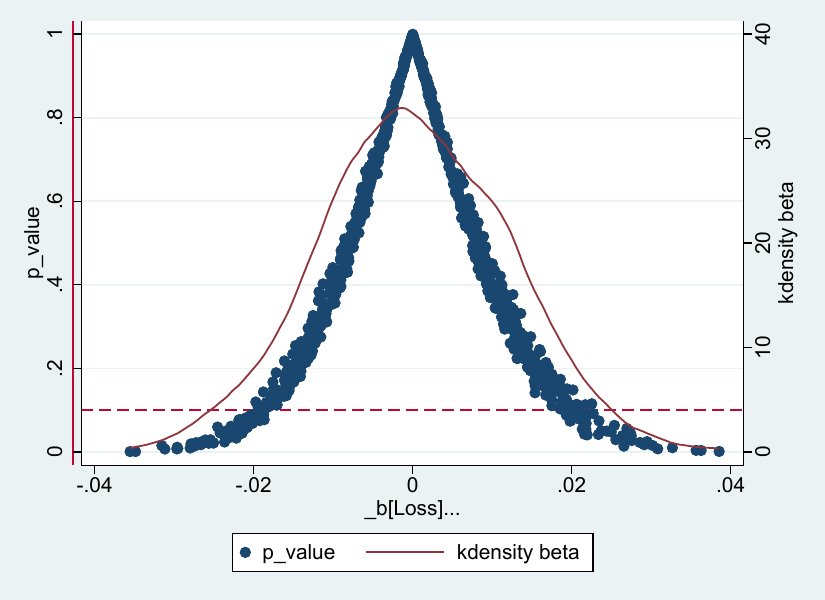}
    \caption{Placebo test (CSR\_Div\_ent).}
    \caption*{\footnotesize\emph{Note:} The x-axis plots coefficients from 1,000 placebo regressions with randomized Loss;dark blue dots are the corresponding $p$-values (left axis), and the red curve is the kernel density (right axis). The vertical solid line marks the baseline estimate $\hat{\beta}=-0.0427$, and the horizontal dashed line indicates $p=0.10$.}
    \label{fig:placebo_entropy}
\end{figure}

Figures~\ref{fig:placebo_gini} and \ref{fig:placebo_entropy} plot the kernel density of the placebo coefficients with corresponding $p$-values (dark blue dots). The vertical solid lines mark the baseline estimates ($-0.0174$ and $-0.0427$), and the horizontal dashed line indicates $p=0.10$. The placebo coefficient distributions are centered near zero, and most $p$-values exceed 0.10, suggesting that unobserved factors are unlikely to generate the documented negative association. We therefore interpret the baseline effect—losses reduce CSR topic diversity—as robust rather than a placebo artifact.

\begin{center}
\textbf{[Insert Figure~\ref{fig:placebo_gini} Here]}
\end{center}

\begin{center}
\textbf{[Insert Figure~\ref{fig:placebo_entropy} Here]}
\end{center}

\subsection{Further Discussion}
The preceding analysis shows that loss-making firms reduce CSR topic diversity under resource constraints, consistent with the slack resources hypothesis and Legitimacy theory. However, when firms face financial distress, their allocation of resources and disclosure strategies are not fixed. They are shaped by the external institutional environment and internal operating characteristics. In different settings, the trade-off between resource constraints and legitimacy maintenance shifts. To delineate the boundary conditions of the main effect, we conduct heterogeneity tests along four dimensions: external oversight (third-party assurance), internal disclosure focus (work-safety content), firm characteristics (large scale), and product-market competition (competitive industries).

\subsubsection{External Oversight: Third-Party Assurance}

\citet{garcia2022assurance} argue that third-party assurance can narrow the ``decoupling'' gap between CSR disclosure and underlying performance. As an external governance mechanism, assurance enhances credibility and transparency, encouraging loss-making firms to broaden disclosure to repair reputation and attract investors. Conversely, firms with stronger CSR disclosure quality are more willing to seek assurance, further bolstering credibility. We therefore interact losses with an indicator for third-party assurance (Loss$\times$Certification), which equals 1 if the CSR report is assured by an external auditor or certification body and 0 otherwise. Columns (1)–(2) of Table~\ref{Heterogeneity1} show coefficients of 0.0882 and 0.2378, significant at the 5\% and 1\% levels, respectively. This indicates that for loss-making firms, third-party assurance is associated with higher CSR topic diversity. Put differently, the negative effect of losses on CSR topic diversity is much weaker when reports are assured and stronger when they are not.

\subsubsection{Disclosure Strategy: Work-Safety Content}

When resources are tight, firms may pivot from broad disclosure to a focused agenda, allocating scarce resources to topics that most directly reflect core responsibilities and managerial capability. Work safety is foundational to legitimacy, given its salience for employee welfare, operational continuity, and regulatory compliance. We therefore interact losses with an indicator for disclosing work-safety content (Loss$\times$WorkSafety), which equals 1 if the CSR report covers work safety and 0 otherwise. Columns (3)–(4) of Table~\ref{Heterogeneity1} show significantly negative interaction coefficients, implying that among loss-making firms, those that disclose work-safety content exhibit lower CSR topic diversity.

At first glance, this may seem counterintuitive because one might expect work-safety disclosure to accompany broader reporting. Our results suggest instead that work-safety disclosure functions as a legitimacy signal. In many real-economy sectors, regulators, employees, and local communities place the greatest emphasis on safety. During loss periods, highlighting safety reveals a shift of limited resources toward a legitimacy-critical topic, which compresses breadth elsewhere in the report.

\begin{center}
\textbf{[Insert Table~\ref{Heterogeneity1} Here]}
\end{center}

\subsubsection{Firm Characteristics: Large-Scale Firms}

Firm size shapes slack resources, disclosure capacity, and reputation constraints. Relative to smaller firms, large firms generally have more mature compliance and reporting processes and can maintain baseline breadth even under financial pressure. They also face stronger scrutiny from investors, media, and regulators, making sharp cutbacks in CSR disclosure less likely. On the other hand, scale can induce more templated reporting and thematic concentration. Using the annual mean of total assets as a cutoff, we classify firms into large and small and estimate the interaction (Loss$\times$LargeScale). Columns (1)–(2) of Table~\ref{Heterogeneity2} report coefficients of $0.0297$ (significant at 5\%) and $0.0732$ (significant at 1\%), indicating that large loss-making firms exhibit higher CSR topic diversity.

Two mechanisms may explain this pattern. First, larger firms benefit from fixed disclosure infrastructures that reduce the marginal cost of maintaining breadth in adverse states. Second, higher external visibility and reputation concerns strengthen managers’ incentives to supply information in downturns to stabilize expectations among investors and regulators.

\subsubsection{Product-Market Competition: Competitive Industries}

Product-market competition is a key external force shaping corporate behavior. In highly competitive markets, survival pressure intensifies, pushing the resource-constraint logic to the extreme. Price wars and share battles compress profit margins for loss-making firms, reinforcing survival-first principles. At the same time, competition acts as an external governance mechanism that reduces slack and agency costs~\citep{giroud2011corporate}. In such environments, expenditures without immediate payoffs are more likely to be cut, and CSR outlays tend to shrink.

We therefore examine competition-based heterogeneity using the industry classification in \citet{YuanChun_XiaoTuSheng_GengChunXiao_ShengYu_2021}, which maps the CSRC 2012 standard into competitive versus non-competitive industries. Columns (3)–(4) of Table~\ref{Heterogeneity2} show significantly negative interaction coefficients of -0.0289 and -0.0633. This indicates that in competitive industries, loss-making firms reduce CSR topic diversity more aggressively. The result underscores competition’s screening role: it aligns resource allocation and disclosure tightly with short-term survival goals, strengthening the explanatory power of the resource-constraint channel while weakening the impetus for broad disclosure rooted in longer-term legitimacy considerations.

\begin{center}
\textbf{[Insert Table~\ref{Heterogeneity2} Here]}
\end{center}

\section{Conclusion and Implications}

Drawing on CSR reports and financial data for Chinese A-share firms from 2006–2023, this paper quantifies CSR topic diversity via LDA and estimates a two-way fixed-effects model to examine how corporate losses affect disclosure structure. We find that losses significantly reduce CSR topic diversity, consistent with the slack resources hypothesis; this result is robust to instrumental variables, propensity score matching, and placebo tests. Mechanism analyses show that media attention, executive compensation incentives, and supervisory board shareholding mitigate the negative impact of losses on CSR topic diversity. Heterogeneity tests further indicate that, among loss-making firms, reports with third-party assurance and reports without work-safety content display relatively higher CSR topic diversity; large firms and firms in less competitive industries also tend to maintain higher diversity.

This study makes several contributions. Methodologically, we shift the evaluation of CSR reporting from intensity to structure by using LDA to uncover latent topic distributions and translating them into comparable diversity metrics. Theoretically, we integrate the logic of resource constraints and Legitimacy theory into a unified framework to explain why managers in distress adopt a ``minimum-cost compliance'' strategy and how external oversight and internal governance expand or narrow the decision space. Substantively, we document the nonfinancial consequences of financial distress along a structural dimension—CSR topic diversity—thereby enriching the notion of CSR disclosure quality.

Our findings yield practical implications for multiple stakeholders. For corporate managers, the key is to avoid treating CSR communication as a fair-weather ``image project'' that can be downsized first when resources tighten. Loss periods are precisely when confidence among investors, employees, and supply-chain partners is most fragile. Substantial contraction in topic coverage sends a strong negative signal that the firm deprioritizes long-term value and the social contract, potentially triggering financing frictions and talent losses. High-powered pay, linking compensation performance to CSR, and increasing supervisory board equity stakes can raise monitoring intensity and help prevent such myopic choices.

For regulators, the focus of CSR oversight should move from formal compliance to substantive disclosure quality. Financial distress is a key trigger for disclosure shrinkage. One-size-fits-all annual requirements are insufficient to curb opportunistic behavior under pressure. A more targeted and dynamic regime is needed: place firms with consecutive losses or deteriorating financial indicators under enhanced supervision; require these firms to explain how financial conditions affect their capacity to fulfill responsibilities; and increase the likelihood of sampling or mandating third-party assurance. Guidance should emphasize the ``structural completeness'' of disclosure—balanced coverage across environmental, social, and governance dimensions—to deter selective emphasis on ``safe'' topics and the concealment of risks in supply chains or human capital. Finally, building a public, searchable database—with indicators such as annual changes in CSR topic diversity—would improve comparability, empower media and the public as external monitors, and strengthen market discipline for comprehensive and transparent CSR communication across the cycle.

For investors and ESG rating agencies, CSR topic diversity offers a complementary signal. A marked post-loss contraction in topic breadth need not be inherently negative, but it reveals the firm’s resource-allocation preferences and shifting strategic priorities under stress. Tracking this shift can help identify early signs of short-termism or weakening governance, thereby improving risk detection and the assessment of long-horizon value.

This study has limitations that suggest avenues for future research. LDA is well suited to capture breadth and structure but less able to assess the factual depth, specificity, or tone of content within each topic. Future work could combine topic modeling with sentiment analysis, readability metrics, and automated fact extraction to build a more multidimensional measure of CSR information quality. Moreover, while we document a macro-level association between losses and disclosure contraction, the internal decision process—how managers trade off topics—is still a ``black box.'' Case studies or executive interviews could help open this box. Finally, our sample is restricted to formal CSR reports. Future research might examine whether disclosures on social media and in news outlets substitute for or complement CSR reports and incorporate the upfront decision of whether to issue a report, providing a more complete view of how financial distress shapes overall transparency.

\section{Tables and Figures}

\begin{table}[H]
\def\sym#1{\ifmmode^{#1}\else\(^{#1}\)\fi}
  \centering
  \caption{Definition of Main Variables}
  \label{tab:variable_description}
  \begin{tabularx}{\textwidth}{>{\raggedright\arraybackslash}p{3cm} >{\raggedright\arraybackslash}p{2.5cm} X}
  \toprule
  Variable & Source & Definition \\
\midrule
CSR\_Div\_gini & Firm Disclosure & The thematic diversity of the firm's CSR report, measured via Latent Dirichlet Allocation (LDA) topic modeling and proxied by the Gini coefficient of the topic distribution. \\
CSR\_Div\_ent  & Firm Disclosure & The thematic diversity of the firm's CSR report, measured via LDA topic modeling and proxied by the entropy of the topic distribution. \\
Loss           & CSMAR           & An indicator equal to 1 if the firm reports a negative net profit in the year, and 0 otherwise. \\
Mt\_positive   & CNRDS           & Total number of positive news articles about the firm in a given year from the CNRDS database. \\
Mt\_neutral    & CNRDS           & Total number of neutral news articles about the firm in a given year from the CNRDS database. \\
Mt\_negative   & CNRDS           & Total number of negative news articles about the firm in a given year from the CNRDS database. \\
Salarytop3     & CSMAR           & Natural logarithm of the total compensation paid to the three highest-paid directors, supervisors, and senior executives in the year. \\
Salarysum      & CSMAR           & Natural logarithm of the total compensation paid to all directors, supervisors, and senior executives in the year. \\
Sshrrat        & CSMAR           & Shareholding ratio of the supervisory board (shares held by all supervisors divided by total shares outstanding). \\
Size           & CSMAR           & Natural logarithm of total assets at fiscal year-end. \\
Lev            & CSMAR           & Total liabilities divided by total assets. \\
Cashflow       & CSMAR           & Net cash flow from operating activities divided by total assets. \\
Fixed          & CSMAR           & Net fixed assets divided by total assets. \\
Top1           & CSMAR           & Proportion of shares held by the largest shareholder (shares held by the largest shareholder divided by total shares outstanding). \\
Board          & CSMAR           & Natural logarithm of the number of board directors. \\
TobinQ         & CSMAR           & (Market value of equity + book value of debt) divided by total assets. \\
\bottomrule
\end{tabularx}
\end{table}

\begin{landscape}
\begin{table}[H]
\centering
\def\sym#1{\ifmmode^{#1}\else\(^{#1}\)\fi}
\caption{Descriptive Statistics} 
\label{tab:descriptive}
\adjustbox{width=0.9\linewidth,center}
{
\begin{tabular}{l*{1}{cccccccc}}
\toprule
                        &           N&        Mean&          SD&         Min&         p25&         p50&         p75&         Max\\
\midrule
CSR\_Div\_gini            &       13,797&      0.4768&      0.2523&      0.0001&      0.2839&      0.5284&      0.6823&      0.9900\\
CSR\_Div\_ent         &       13,797&      1.0089&      0.5913&      0.0009&      0.5704&      1.0186&      1.4381&      4.6052\\
Loss                    &       13,797&      0.1074&      0.3097&      0.0000&      0.0000&      0.0000&      0.0000&      1.0000\\
Mt\_positive            &       12,888&    214.4731&    525.3398&      0.0000&     42.0000&     87.0000&    198.0000&    15475\\
Mt\_neutral            &       12,888&    117.7293&    340.5424&      0.0000&     21.0000&     42.0000&     93.0000&     11123 \\
Mt\_negative            &       12,888&    150.1034&    463.6001&      0.0000&     23.0000&     49.0000&    119.0000&     22873\\
Salarytop3              &       13,783&     14.9056&      0.9864&      0.0000&     14.4079&     14.8675&     15.4065&     18.5840\\
Salarysum               &       13,777&     15.7354&      0.9794&      0.0000&     15.2026&     15.7193&     16.2592&     18.9301\\
Sshrrat                 &       11,886&      0.1410&      1.0221&      0.0000&      0.0000&      0.0000&      0.0014&     48.4637\\
Size                    &       13,797&     23.3738&      1.7507&     18.2659&     22.1375&     23.1312&     24.2874&     31.4309\\
Lev                     &       13,797&      0.4934&      0.2150&      0.0080&      0.3321&      0.4970&      0.6479&      2.2901\\
Cashflow                &       13,797&      0.0525&      0.0756&     -0.5173&      0.0138&      0.0508&      0.0921&      0.7255\\
Fixed                   &       13,797&      0.2083&      0.1791&      0.0000&      0.0658&      0.1628&      0.3126&      0.9542\\
Top1                    &       13,797&      0.3584&      0.1622&      0.0223&      0.2292&      0.3401&      0.4773&      0.8999\\
Board                   &       13,797&      2.1861&      0.2237&      1.3863&      2.0794&      2.1972&      2.1972&      3.0445\\
TobinQ                  &       13,797&      1.8183&      1.4208&      0.6085&      1.0764&      1.3825&      1.9922&     29.1670\\
\bottomrule
\multicolumn{9}{l}{\footnotesize \textbf{Notes: }This table presents descriptive statistics of the main variables used in this paper.}\\
\end{tabular}
}
\end{table}

\end{landscape}

\begin{table}[H]\centering
\def\sym#1{\ifmmode^{#1}\else\(^{#1}\)\fi}
\caption{Baseline Regression}
\label{tab:CSR_FinancialDistress}
\begin{adjustbox}{width=\textwidth,center}
\begin{threeparttable}
\begin{tabular}{l*{4}{c}}
\toprule
                         &\multicolumn{1}{c}{(1)}&\multicolumn{1}{c}{(2)}&\multicolumn{1}{c}{(3)}&\multicolumn{1}{c}{(4)}\\
                         &\multicolumn{1}{c}{CSR\_Div\_gini}&\multicolumn{1}{c}{CSR\_Div\_gini}&\multicolumn{1}{c}{CSR\_Div\_ent}&\multicolumn{1}{c}{CSR\_Div\_ent}\\
\midrule
Loss                     &     -0.0164\sym{***}&     -0.0174\sym{***}&     -0.0424\sym{***}&     -0.0427\sym{***}\\
                         &    (0.0062)         &    (0.0063)         &    (0.0150)         &    (0.0154)         \\
\addlinespace
Size                     &                     &     -0.0085         &                     &     -0.0134         \\
                         &                     &    (0.0084)         &                     &    (0.0190)         \\
\addlinespace
Lev                      &                     &      0.0176         &                     &      0.0151         \\
                         &                     &    (0.0316)         &                     &    (0.0723)         \\
\addlinespace
Cashflow                 &                     &      0.0415         &                     &      0.1201\sym{*}  \\
                         &                     &    (0.0283)         &                     &    (0.0665)         \\
\addlinespace
Fixed                   &                     &      0.0491         &                     &      0.1281         \\
                         &                     &    (0.0354)         &                     &    (0.0811)         \\
\addlinespace
Top1                     &                     &      0.0382         &                     &      0.0804         \\
                         &                     &    (0.0515)         &                     &    (0.1145)         \\
\addlinespace
Board                    &                     &     -0.0004         &                     &      0.0005         \\
                         &                     &    (0.0203)         &                     &    (0.0487)         \\
\addlinespace
TobinQ                   &                     &      0.0010         &                     &      0.0010         \\
                         &                     &    (0.0024)         &                     &    (0.0057)         \\
\addlinespace
Constant                 &      0.4490\sym{***}&      0.6113\sym{***}&      0.9391\sym{***}&      1.1808\sym{***}\\
                         &    (0.0006)         &    (0.1968)         &    (0.0014)         &    (0.4470)         \\
\midrule
Individual FE & YES & YES & YES & YES \\
Year FE       & YES & YES & YES & YES  \\
Observations             &       10,849         &       10,849         &       10,849         &       10,849         \\
Adjusted $R^2$             &       0.703         &       0.703         &       0.658         &       0.659        \\
\bottomrule
\end{tabular}
\begin{tablenotes}[para]
  \footnotesize
      \item \textbf{Notes:} 
      This table reports the results of baseline regression. \sym{*}, \sym{*}\sym{*}, and \sym{*}\sym{*}\sym{*} indicate significance at 10\%, 5\%, and 1\% levels, respectively. The standard errors clustered at the firm level are presented in parentheses.
\end{tablenotes}
\end{threeparttable}
\end{adjustbox}
\end{table}
\begin{landscape}
\begin{table}[htbp]\centering
\caption{Monitoring Role of Media Attention}
\label{tab:news_moderation}
\def\sym#1{\ifmmode^{#1}\else\(^{#1}\)\fi}
\begin{adjustbox}{width=0.8\linewidth,center}
\begin{threeparttable}
\begin{tabular}{l*{6}{c}}
\toprule
            &\multicolumn{2}{c}{Positive News}          &\multicolumn{2}{c}{Neutral News}           &\multicolumn{2}{c}{Negative News}          \\\cmidrule(lr){2-3}\cmidrule(lr){4-5}\cmidrule(lr){6-7}
            &\multicolumn{1}{c}{(1)}&\multicolumn{1}{c}{(2)}&\multicolumn{1}{c}{(3)}&\multicolumn{1}{c}{(4)}&\multicolumn{1}{c}{(5)}&\multicolumn{1}{c}{(6)}\\
            &\multicolumn{1}{c}{CSR\_Div\_gini}&\multicolumn{1}{c}{CSR\_Div\_ent}&\multicolumn{1}{c}{CSR\_Div\_gini}&\multicolumn{1}{c}{CSR\_Div\_ent}&\multicolumn{1}{c}{CSR\_Div\_gini}&\multicolumn{1}{c}{CSR\_Div\_ent}\\
\midrule
Loss      &    -0.02233\sym{***}&    -0.05094\sym{***}&    -0.02170\sym{***}&    -0.04950\sym{***}&    -0.02212\sym{***}&    -0.05190\sym{***}\\
            &   (0.00690)         &   (0.01693)         &   (0.00664)         &   (0.01626)         &   (0.00669)         &   (0.01643)         \\
\addlinespace
Mt          &    -0.00001         &    -0.00002         &      -0.00001       &   -0.00002         &     -0.00001\sym{*}     &     -0.00003\sym{*}     \\
            &   (0.00001)         &   (0.00002)         &    (0.00001)        &   (0.00002)         &      (0.00001)       &          (0.00002)    \\
\addlinespace
Loss$×$Mt&     0.00002\sym{*}  &     0.00004      &   0.00003\sym{**}  &    0.00005           &     0.00003\sym{***}   &   0.00005\sym{**}          \\
            &   (0.00001)         &   (0.00003)         &      (0.00001)    &     (0.00004)        &          (0.00001)           &     (0.00002)           \\
\addlinespace
Size        &    -0.00753         &    -0.01128         &    -0.00745         &    -0.01125         &    -0.00733         &    -0.01109         \\
            &   (0.00849)         &   (0.01944)         &   (0.00851)         &   (0.01947)         &   (0.00849)         &   (0.01940)         \\
\addlinespace
Lev         &     0.01636         &     0.01807         &     0.01615         &     0.01769         &     0.01620         &     0.01807         \\
            &   (0.03173)         &   (0.07343)         &   (0.03171)         &   (0.07340)         &   (0.03173)         &   (0.07343)         \\
\addlinespace
Cashflow    &     0.08817\sym{***}&     0.24273\sym{***}&     0.08781\sym{***}&     0.24164\sym{***}&     0.08782\sym{***}&     0.24214\sym{***}\\
            &   (0.02979)         &   (0.07162)         &   (0.02981)         &   (0.07166)         &   (0.02981)         &   (0.07161)         \\
\addlinespace
Fixed       &     0.05028         &     0.12924         &     0.05035         &     0.12928         &     0.05072         &     0.13034         \\
            &   (0.03541)         &   (0.08128)         &   (0.03541)         &   (0.08127)         &   (0.03537)         &   (0.08120)         \\
\addlinespace
Top1        &     0.02934         &     0.05505         &     0.02989         &     0.05616         &     0.02815         &     0.05265         \\
            &   (0.04953)         &   (0.11209)         &   (0.04956)         &   (0.11215)         &   (0.04939)         &   (0.11184)         \\
\addlinespace
Board       &     0.00210         &     0.00851         &     0.00244         &     0.00916         &     0.00230         &     0.00899         \\
            &   (0.02106)         &   (0.05122)         &   (0.02107)         &   (0.05124)         &   (0.02107)         &   (0.05124)         \\
\addlinespace
TobinQ      &     0.00093         &     0.00086         &     0.00089         &     0.00076         &     0.00089         &     0.00074         \\
            &   (0.00248)         &   (0.00581)         &   (0.00248)         &   (0.00582)         &   (0.00248)         &   (0.00581)         \\
\midrule
Individual FE&         YES         &         YES         &         YES            &     YES                &      YES               &            YES         \\
Year FE     &         YES         &         YES         &             YES        &          YES           &          YES           &          YES           \\
Observations&      10,048         &      10,048         &      10,048         &      10,048         &      10,048         &      10,048         \\
Adjusted R$^{2}$&      0.717         &      0.674         &      0.717         &      0.674         &      0.717         &      0.674         \\
\bottomrule
\end{tabular}
\begin{tablenotes}[para]
  \footnotesize
      \item \textbf{Notes:} 
      This table reports regression results on the moderating effect of management compensation.\sym{*}, \sym{*}\sym{*}, and \sym{*}\sym{*}\sym{*} indicate significance at 10\%, 5\%, and 1\% levels, respectively. The standard errors clustered at the firm level are presented in parentheses.
\end{tablenotes}
\end{threeparttable}
\end{adjustbox}
\end{table}

\end{landscape}

\begin{table}[H]\centering
\def\sym#1{\ifmmode^{#1}\else\(^{#1}\)\fi}
\caption{Incentive effect of management compensation}
\label{tab:pay_moderation}
\begin{adjustbox}{width=\textwidth,center}
\begin{threeparttable}
\begin{tabular}{l*{4}{c}}
\toprule
                         &\multicolumn{1}{c}{(1)}&\multicolumn{1}{c}{(2)}&\multicolumn{1}{c}{(3)}&\multicolumn{1}{c}{(4)}\\
                         &\multicolumn{1}{c}{CSR\_Div\_gini}&\multicolumn{1}{c}{CSR\_Div\_ent}&\multicolumn{1}{c}{CSR\_Div\_gini}&\multicolumn{1}{c}{CSR\_Div\_ent}\\
\midrule

Loss                   &     -0.2527\sym{**} &     -0.5381\sym{**} &     -0.1905\sym{*}  &     -0.4490\sym{*}  \\
                         &    (0.1046)         &    (0.2433)         &    (0.1016)         &    (0.2575)         \\
\addlinespace
Salarytop3&      0.0007         &      0.0040         &                     &                     \\
                         &    (0.0034)         &    (0.0078)         &                     &                     \\
\addlinespace
Loss$\times$Salarytop3&      0.0161\sym{**} &      0.0340\sym{**} &                     &                     \\
                         &    (0.0072)         &    (0.0167)         &                     &                     \\
salarysum&                     &                     &     -0.0031         &     -0.0038         \\
                         &                     &                     &    (0.0041)         &    (0.0094)         \\
\addlinespace
 Loss$\times$Salarysum&                     &                     &      0.0112\sym{*}  &      0.0264         \\
                         &                     &                     &    (0.0066)         &    (0.0168)         \\
\addlinespace
\addlinespace
Size                     &     -0.0084         &     -0.0140         &     -0.0074         &     -0.0120         \\
                         &    (0.0084)         &    (0.0190)         &    (0.0084)         &    (0.0189)         \\
\addlinespace
Lev                      &      0.0187         &      0.0181         &      0.0177         &      0.0164         \\
                         &    (0.0316)         &    (0.0724)         &    (0.0316)         &    (0.0725)         \\
\addlinespace
Cashflow                 &      0.0411         &      0.1187\sym{*}  &      0.0414         &      0.1192\sym{*}  \\
                         &    (0.0283)         &    (0.0665)         &    (0.0283)         &    (0.0667)         \\
\addlinespace
Fixed                    &      0.0522         &      0.1352\sym{*}  &      0.0484         &      0.1272         \\
                         &    (0.0355)         &    (0.0813)         &    (0.0354)         &    (0.0815)         \\
\addlinespace
Top1                     &      0.0374         &      0.0793         &      0.0365         &      0.0775         \\
                         &    (0.0516)         &    (0.1148)         &    (0.0516)         &    (0.1149)         \\
\addlinespace
Board                    &     -0.0014         &     -0.0017         &     -0.0000         &      0.0002         \\
                         &    (0.0203)         &    (0.0485)         &    (0.0204)         &    (0.0490)         \\
\addlinespace
TobinQ                   &      0.0011         &      0.0012         &      0.0011         &      0.0012         \\
                         &    (0.0024)         &    (0.0057)         &    (0.0024)         &    (0.0057)         \\
\addlinespace
Constant                 &      0.6013\sym{***}&      1.1385\sym{**} &      0.6353\sym{***}&      1.2093\sym{***}\\
                         &    (0.1992)         &    (0.4545)         &    (0.2012)         &    (0.4582)         \\
\midrule
Individual FE&         YES         &         YES         &         YES            &     YES                  \\
Year FE     &         YES         &         YES         &             YES        &          YES                    \\
Observations             &       10,838         &       10,838         &       10,834         &       10,834         \\
Adjusted $R^2$             &       0.703         &       0.658         &       0.703         &       0.658         \\
\bottomrule
\end{tabular}
\begin{tablenotes}[para]
  \footnotesize
  \item \textbf{Notes:} This table examines the moderating effects of executive compensation on CSR topic diversity. Columns (1)–(2) use \texttt{Salarytop3} as the compensation measure (top-three executives), while columns (3)–(4) use \texttt{salarysum} (aggregate management pay). 
  Standard errors are clustered at the firm level and reported in parentheses. 
  \sym{*}, \sym{**}, and \sym{***} indicate significance at the 10\%, 5\%, and 1\% levels, respectively.
\end{tablenotes}
\end{threeparttable}
\end{adjustbox}
\end{table}

\begin{table}[H]\centering
\def\sym#1{\ifmmode^{#1}\else\(^{#1}\)\fi}
\caption{The supervisory role of the board of supervisors}
\label{tab:supervisory_share_moderation}
\begin{threeparttable}

\begin{tabular}{l*{2}{c}}
\toprule
                         &\multicolumn{1}{c}{(1)}&\multicolumn{1}{c}{(2)}\\
                         &\multicolumn{1}{c}{CSR\_Div\_gini}&\multicolumn{1}{c}{CSR\_Div\_ent}\\
\midrule
Loss                     &     -0.0161\sym{**}  &     -0.0428\sym{***}\\
                         &    (0.0067)          &    (0.0156)         \\
\addlinespace
SshrRat                  &     -0.0036\sym{**}  &     -0.0093\sym{**} \\
                         &    (0.0018)          &    (0.0040)         \\
\addlinespace
Loss$\times$SshrRat      &      0.0041\sym{**}  &      0.0128\sym{***}\\
                         &    (0.0021)          &    (0.0046)         \\
\addlinespace
Size                     &     -0.0089          &     -0.0154         \\
                         &    (0.0088)          &    (0.0199)         \\
\addlinespace
Lev                      &      0.0293          &      0.0567         \\
                         &    (0.0336)          &    (0.0755)         \\
\addlinespace
Cashflow                 &      0.0405          &      0.0952         \\
                         &    (0.0294)          &    (0.0671)         \\
\addlinespace
Fixed                    &      0.0537          &      0.1194         \\
                         &    (0.0364)          &    (0.0831)         \\
\addlinespace
Top1                     &      0.0210          &      0.0485         \\
                         &    (0.0534)          &    (0.1191)         \\
\addlinespace
Board                    &     -0.0024          &      0.0003         \\
                         &    (0.0213)          &    (0.0507)         \\
\addlinespace
TobinQ                   &      0.0010          &      0.0005         \\
                         &    (0.0028)          &    (0.0065)         \\
\addlinespace
\_cons                   &      0.6105\sym{***} &      1.1862\sym{**} \\
                         &    (0.2054)          &    (0.4669)         \\
\midrule
Individual FE            & \multicolumn{1}{c}{YES} & \multicolumn{1}{c}{YES} \\
Year FE                  & \multicolumn{1}{c}{YES} & \multicolumn{1}{c}{YES} \\
Observations             &        9,333          &        9,333         \\
Adjusted R$^2$          &        0.716          &        0.672         \\
\bottomrule
\end{tabular}

\begin{tablenotes}[para]
  \footnotesize
  \item \textbf{Notes:} 
  This table reports regression results on the supervisory role of the board of supervisors. Standard errors are clustered at the firm level and reported in parentheses. \sym{*}, \sym{**}, and \sym{***} indicate significance at the 10\%, 5\%, and 1\% levels, respectively.
\end{tablenotes}

\end{threeparttable}
\end{table}

\begin{table}[H]\centering
\def\sym#1{\ifmmode^{#1}\else\(^{#1}\)\fi}
\begin{adjustbox}{width=\linewidth,center}
\begin{threeparttable}

\caption{Winsorization and Exclusion of Special Observations}
\label{tab:robustness1.1}
\begin{tabular}{l*{4}{c}}
\toprule
                         &\multicolumn{1}{c}{(1)}&\multicolumn{1}{c}{(2)}&\multicolumn{1}{c}{(3)}&\multicolumn{1}{c}{(4)}\\
                         &\multicolumn{1}{c}{CSR\_Div\_gini}&\multicolumn{1}{c}{CSR\_Div\_ent}&\multicolumn{1}{c}{CSR\_Div\_gini}&\multicolumn{1}{c}{CSR\_Div\_ent}\\
\midrule
Loss                     &     -0.0174\sym{***}&     -0.0427\sym{***}&     -0.0166\sym{***}&     -0.0414\sym{***}\\
                         &    (0.0063)         &    (0.0154)         &    (0.0064)         &    (0.0158)         \\
\addlinespace
Size                     &     -0.0085         &     -0.0134         &     -0.0071         &     -0.0088         \\
                         &    (0.0084)         &    (0.0190)         &    (0.0090)         &    (0.0205)         \\
\addlinespace
Lev                      &      0.0176         &      0.0151         &      0.0187         &      0.0255         \\
                         &    (0.0316)         &    (0.0723)         &    (0.0326)         &    (0.0749)         \\
\addlinespace
Cashflow                 &      0.0415         &      0.1201\sym{*}  &      0.0956\sym{***}&      0.2597\sym{***}\\
                         &    (0.0283)         &    (0.0665)         &    (0.0304)         &    (0.0730)         \\
\addlinespace
Fixed                   &      0.0491         &      0.1281         &      0.0532         &      0.1300         \\
                         &    (0.0354)         &    (0.0811)         &    (0.0357)         &    (0.0817)         \\
\addlinespace
Top1                     &      0.0382         &      0.0804         &      0.0371         &      0.0748         \\
                         &    (0.0515)         &    (0.1145)         &    (0.0506)         &    (0.1141)         \\
\addlinespace
Board                    &     -0.0004         &      0.0005         &     -0.0000         &      0.0028         \\
                         &    (0.0203)         &    (0.0487)         &    (0.0215)         &    (0.0521)         \\
\addlinespace
TobinQ                   &      0.0010         &      0.0010         &      0.0012         &      0.0015         \\
                         &    (0.0024)         &    (0.0057)         &    (0.0025)         &    (0.0058)         \\
\addlinespace
Constant                 &      0.6113\sym{***}&      1.1808\sym{***}&      0.5711\sym{***}&      1.0536\sym{**} \\
                         &    (0.1968)         &    (0.4470)         &    (0.2109)         &    (0.4799)         \\
\midrule
Individual FE & YES  & YES  & YES  & YES  \\
Year FE       & YES  & YES  & YES  & YES  \\
Observations             &       10,849         &       10,849         &        9,971         &        9,971         \\
Adjusted R\textsuperscript{2} &       0.703         &       0.659         &       0.715         &       0.671         \\
\bottomrule
\end{tabular}

\begin{tablenotes}[para]
  \footnotesize
 \item \textbf{Notes:} This table reports regression results after winsorization and the exclusion of special observations.
  Columns (1)–(2) report estimates after winsorizing continuous variables at the 1\% level on both tails. 
 Columns (3)–(4) report estimates after excluding firms designated ST or *ST and those in the financial industry.
  Standard errors are clustered at the firm level and reported in parentheses. 
  \sym{*}, \sym{**}, and \sym{***} indicate significance at the 10\%, 5\%, and 1\% levels, respectively.
\end{tablenotes}

\end{threeparttable}
\end{adjustbox}
\end{table}

\begin{table}[H]\centering
\def\sym#1{\ifmmode^{#1}\else\(^{#1}\)\fi}
\begin{adjustbox}{width=\linewidth,center}
\begin{threeparttable}

\caption{Add an Extra Control and Replace the Explanatory Variable}
\label{tab:robustness1}

\begin{tabular}{l*{4}{c}}
\toprule
                         &\multicolumn{1}{c}{(1)}&\multicolumn{1}{c}{(2)}&\multicolumn{1}{c}{(3)}&\multicolumn{1}{c}{(4)}\\
                         &\multicolumn{1}{c}{CSR\_Div\_gini}&\multicolumn{1}{c}{CSR\_Div\_ent}&\multicolumn{1}{c}{CSR\_Div\_gini}&\multicolumn{1}{c}{CSR\_Div\_ent}\\
\midrule
Loss                     &     -0.0175\sym{***}&     -0.0428\sym{***}&                     &                     \\
                         &    (0.0063)         &    (0.0154)         &                     &                     \\
Roa                      &                     &                     &      0.0708\sym{*}  &      0.2256\sym{**} \\
                         &                     &                     &    (0.0404)         &    (0.0923)         \\
Size                     &     -0.0080         &     -0.0118         &     -0.0084         &     -0.0145         \\
                         &    (0.0084)         &    (0.0190)         &    (0.0085)         &    (0.0191)         \\
Lev                      &      0.0171         &      0.0133         &      0.0202         &      0.0323         \\
                         &    (0.0317)         &    (0.0725)         &    (0.0324)         &    (0.0735)         \\
Cashflow                 &      0.0424         &      0.1229\sym{*}  &      0.0321         &      0.0850         \\
                         &    (0.0283)         &    (0.0664)         &    (0.0291)         &    (0.0676)         \\
Fixed                    &      0.0489         &      0.1277         &      0.0514         &      0.1381\sym{*}  \\
                         &    (0.0354)         &    (0.0811)         &    (0.0355)         &    (0.0814)         \\
Top1                     &      0.0392         &      0.0837         &      0.0385         &      0.0789         \\
                         &    (0.0514)         &    (0.1144)         &    (0.0514)         &    (0.1143)         \\
Board                    &     -0.0006         &     -0.0002         &     -0.0013         &     -0.0020         \\
                         &    (0.0203)         &    (0.0486)         &    (0.0203)         &    (0.0488)         \\
TobinQ                   &      0.0011         &      0.0012         &      0.0007         &     -0.0001         \\
                         &    (0.0024)         &    (0.0057)         &    (0.0025)         &    (0.0058)         \\
GRI                      &     -0.0066         &     -0.0217         &                     &                     \\
                         &    (0.0082)         &    (0.0193)         &                     &                     \\
\_cons                   &      0.6019\sym{***}&      1.1496\sym{**} &      0.6063\sym{***}&      1.1906\sym{***}\\
                         &    (0.1964)         &    (0.4466)         &    (0.1984)         &    (0.4501)         \\
\midrule
Individual FE & YES  & YES  & YES  & YES  \\
Year FE       & YES  & YES  & YES  & YES  \\
Observations  & 10,849 & 10,849 & 10,849 & 10,849 \\
Adjusted $R^2$& 0.703  & 0.659  & 0.703  & 0.659  \\
\bottomrule
\end{tabular}

\begin{tablenotes}[para]
  \footnotesize
 \item \textbf{Notes:} This table presents a subset of robustness analyses. 
  Columns (1)--(2) augment the baseline with an extra control \texttt{GRI}, which equals 1 if the firm's CSR report complies with the Global Reporting Initiative (GRI) standard and 0 otherwise. 
  Columns (3)--(4) replace the key explanatory variable with Roa.  
  Standard errors are clustered at the firm level and reported in parentheses. 
  \sym{*}, \sym{**}, and \sym{***} indicate significance at the 10\%, 5\%, and 1\% levels, respectively.
\end{tablenotes}

\end{threeparttable}
\end{adjustbox}
\end{table}

\begin{landscape}
\begin{table}[H]\centering
\def\sym#1{\ifmmode^{#1}\else\(^{#1}\)\fi}
\begin{adjustbox}{width=\linewidth,center}
\begin{threeparttable}
\caption{The results of two-stage least squares.}
\label{tab:2sls_results}
\begin{tabular}{l*{9}{c}}
\toprule
                    &\multicolumn{3}{c}{$IV_1$}                                  &\multicolumn{3}{c}{$IV_2$}                                  &\multicolumn{3}{c}{$IV_3$}                                  \\\cmidrule(lr){2-4}\cmidrule(lr){5-7}\cmidrule(lr){8-10}
                    &\multicolumn{1}{c}{(1)}&\multicolumn{1}{c}{(2)}&\multicolumn{1}{c}{(3)}&\multicolumn{1}{c}{(4)}&\multicolumn{1}{c}{(5)}&\multicolumn{1}{c}{(6)}&\multicolumn{1}{c}{(7)}&\multicolumn{1}{c}{(8)}&\multicolumn{1}{c}{(9)}\\
                    &\multicolumn{1}{c}{Loss}&\multicolumn{1}{c}{CSR\_Div\_gini}&\multicolumn{1}{c}{CSR\_Div\_ent}&\multicolumn{1}{c}{Loss}&\multicolumn{1}{c}{CSR\_Div\_gini }&\multicolumn{1}{c}{CSR\_Div\_ent}&\multicolumn{1}{c}{Loss}&\multicolumn{1}{c}{CSR\_Div\_gini}&\multicolumn{1}{c}{CSR\_Div\_ent}\\
\midrule
$IV_1$                &    -3.1596\sym{***} &                     &                     &                     &                     &                     &                     &                     &                     \\
                    &    (0.2180)         &                     &                     &                     &                     &                     &                     &                     &                     \\
$IV_2$               &                     &                     &                     &  -1.4616\sym{***}   &                     &                     &                     &                     &                     \\
                    &                     &                     &                     &    (0.1113)         &                     &                     &                     &                     &                     \\
$IV_3$               &                     &                     &                     &                     &                     &                     &   -1.0089\sym{***}  &                     &                     \\
                    &                     &                     &                     &                     &                     &                     &      (0.0738)       &                     &                     \\
Loss                &                     &     -0.2919\sym{***}&     -0.6516\sym{***}&                     &     -0.2869\sym{***}&     -0.6683\sym{***}&                     &     -0.2886\sym{***}&     -0.6629\sym{***}\\
                    &                     &    (0.0411)         &    (0.0916)         &                     &    (0.0414)         &    (0.0960)         &                     &    (0.0411)         &    (0.0942)         \\
Size                &     -0.0487\sym{***}&     -0.0316\sym{***}&     -0.0658\sym{***}&     -0.0488\sym{***}&     -0.0312\sym{***}&     -0.0671\sym{***}&     -0.0485\sym{***}&     -0.0313\sym{***}&     -0.0667\sym{***}\\
                    &    (0.0129)         &    (0.0098)         &    (0.0222)         &    (0.0130)         &    (0.0098)         &    (0.0224)         &    (0.0130)         &    (0.0098)         &    (0.0223)         \\
Lev                 &      0.3469\sym{***}&      0.2033\sym{***}&      0.4416\sym{***}&      0.3589\sym{***}&      0.2001\sym{***}&      0.4521\sym{***}&      0.3526\sym{***}&      0.2011\sym{***}&      0.4487\sym{***}\\
                    &    (0.0537)         &    (0.0416)         &    (0.0928)         &    (0.0542)         &    (0.0418)         &    (0.0945)         &    (0.0541)         &    (0.0417)         &    (0.0938)         \\
Cashflow            &     -0.0665         &     -0.0481         &     -0.0998         &     -0.0835         &     -0.0464         &     -0.1054         &     -0.0756         &     -0.0469         &     -0.1036         \\
                    &    (0.0583)         &    (0.0359)         &    (0.0822)         &    (0.0585)         &    (0.0359)         &    (0.0832)         &    (0.0583)         &    (0.0359)         &    (0.0829)         \\
Fixed               &      0.1025         &      0.1078\sym{**} &      0.2406\sym{**} &      0.1108\sym{*}  &      0.1069\sym{**} &      0.2436\sym{**} &      0.1075\sym{*}  &      0.1071\sym{**} &      0.2426\sym{**} \\
                    &    (0.0652)         &    (0.0423)         &    (0.0971)         &    (0.0652)         &    (0.0422)         &    (0.0978)         &    (0.0652)         &    (0.0422)         &    (0.0976)         \\
Top1                &     -0.1038         &     -0.0179         &     -0.0350         &     -0.1066         &     -0.0170         &     -0.0379         &     -0.1051         &     -0.0173         &     -0.0370         \\
                    &    (0.0662)         &    (0.0597)         &    (0.1330)         &    (0.0664)         &    (0.0595)         &    (0.1333)         &    (0.0663)         &    (0.0596)         &    (0.1332)         \\
Board               &     0.0208          &     -0.0065         &     -0.0116         &     0.0197          &     -0.0066         &     -0.0114         &     0.0202          &     -0.0066         &     -0.0115         \\
                    &    (0.0334)         &    (0.0238)         &    (0.0558)         &    (0.0334)         &    (0.0237)         &    (0.0560)         &    (0.0334)         &    (0.0237)         &    (0.0560)         \\
TobinQ              &     -0.0059         &     -0.0035         &     -0.0086         &     -0.0068         &     -0.0035         &     -0.0089         &     -0.0064         &     -0.0035         &     -0.0088         \\
                    &    (0.0042)         &    (0.0024)         &    (0.0057)         &    (0.0042)         &    (0.0025)         &    (0.0057)         &    (0.0042)         &    (0.0025)         &    (0.0057)         \\
\midrule
Individual FE & YES & YES & YES & YES & YES & YES & YES & YES & YES \\
Year FE       & YES & YES & YES & YES & YES & YES & YES & YES & YES \\
Observations        &        8,896        &        8,896        &        8,896        &        8,896        &        8,896        &        8,896        &        8,896        &        8,896        &        8,896        \\
Number of firms     &       1,153         &       1,153         &       1,153         &       1,153         &       1,153         &       1,153         &       1,153         &       1,153         &       1,153         \\
KP rk LM statistic  & 121.727\sym{***}    &                     &                    & 123.738\sym{***}    &                     &                    & 123.506\sym{***}    &                     &                    \\
KP rk Wald F statistic & 210.099          &                     &                    & 172.512             &                     &                    & 187.061             &                     &                    \\
\bottomrule 
\end{tabular}

\begin{tablenotes}[para]
  \footnotesize
  \item \textbf{Notes:} This table reports two-stage least squares estimates treating Loss as endogenous. Columns (1), (4), and (7) present the first-stage regressions with Loss as the dependent variable; the reported coefficients are on the corresponding instrument. Columns (2)--(3), (5)--(6), and (8)--(9) report the second-stage results for CSR topic diversity measured by a Gini-based concentration index and an entropy index .Standard errors are clustered at the firm level and reported in parentheses. The Kleibergen--Paap rk LM statistic tests underidentification; the Kleibergen--Paap rk Wald \(F\) statistic assesses weak identification. \sym{*}, \sym{**}, and \sym{***} denote significance at the 10\%, 5\%, and 1\% levels, respectively.
\end{tablenotes}
\end{threeparttable}
\end{adjustbox}
\end{table}
\end{landscape}

\begin{table}[H]\centering
\def\sym#1{\ifmmode^{#1}\else\(^{#1}\)\fi}
\caption{Propensity score matching}
\label{tab:robustness_psm}
\begin{threeparttable}

\begin{tabular}{l*{2}{c}}
\toprule
                         &\multicolumn{1}{c}{(1)}&\multicolumn{1}{c}{(2)}\\
                         &\multicolumn{1}{c}{CSR\_Div\_gini}&\multicolumn{1}{c}{CSR\_Div\_ent}\\
\midrule
Loss                     &     -0.0241\sym{*}  &     -0.0497\sym{*}  \\
                         &    (0.0123)         &    (0.0270)         \\
\addlinespace
Size                     &     -0.0156         &      0.0160         \\
                         &    (0.0179)         &    (0.0377)         \\
\addlinespace
Lev                      &     -0.0117         &     -0.1048         \\
                         &    (0.0576)         &    (0.1202)         \\
\addlinespace
Cashflow                 &     -0.0111         &      0.0691         \\
                         &    (0.0681)         &    (0.1499)         \\
\addlinespace
Fixed                    &      0.0030         &      0.1156         \\
                         &    (0.0693)         &    (0.1674)         \\
\addlinespace
Top1                     &      0.0008         &      0.0471         \\
                         &    (0.0953)         &    (0.2460)         \\
\addlinespace
Board                    &     -0.0096         &      0.0632         \\
                         &    (0.0495)         &    (0.1014)         \\
\addlinespace
TobinQ                   &      0.0059         &      0.0050         \\
                         &    (0.0044)         &    (0.0172)         \\
\addlinespace
\_cons                   &      0.8429\sym{**} &      0.4843         \\
                         &    (0.4032)         &    (0.8849)         \\
\midrule
Individual FE            & YES & YES \\
Year FE                  & YES & YES \\
Observations             & 1,377 & 1,460 \\
Adjusted $R^2$           & 0.739 & 0.679 \\
\bottomrule
\end{tabular}

\begin{tablenotes}[para]
  \footnotesize
  \item \textbf{Notes:} This table reports regression results after propensity score matching. \sym{*}, \sym{**}, and \sym{***} indicate significance at the 10\%, 5\%, and 1\% levels, respectively. Standard errors are clustered at the firm level and reported in parentheses. 
\end{tablenotes}

\end{threeparttable}
\end{table}

\begin{table}[H]\centering
\def\sym#1{\ifmmode^{#1}\else\(^{#1}\)\fi}
\caption{Heterogeneity of Certification and WorkSafety}
\label{Heterogeneity1}
\begin{adjustbox}{width=\textwidth,center}
\begin{threeparttable}
\begin{tabular}{l*{4}{c}}
\toprule
                         &\multicolumn{1}{c}{(1)}&\multicolumn{1}{c}{(2)}&\multicolumn{1}{c}{(3)}&\multicolumn{1}{c}{(4)}\\
                         &\multicolumn{1}{c}{CSR\_Div\_gini}&\multicolumn{1}{c}{CSR\_Div\_ent}&\multicolumn{1}{c}{CSR\_Div\_gini}&\multicolumn{1}{c}{CSR\_Div\_ent}\\
\midrule
Loss                     &     -0.0192\sym{***}&     -0.0473\sym{***}&      0.0061         &      0.0120         \\
                         &    (0.0064)         &    (0.0154)         &    (0.0141)         &    (0.0304)         \\
\addlinespace
Certification            &     -0.0296         &     -0.0870\sym{**} &                     &                     \\
                         &    (0.0181)         &    (0.0403)         &                     &                     \\
\addlinespace
Loss$\times$Certification&      0.0882\sym{**} &      0.2378\sym{***}&                     &                     \\
                         &    (0.0380)         &    (0.0911)         &                     &                     \\
\addlinespace
WorkSafety               &                     &                     &      0.0011         &      0.0020         \\
                         &                     &                     &    (0.0063)         &    (0.0140)         \\
\addlinespace
Loss$\times$WorkSafety   &                     &                     &     -0.0274\sym{*}  &     -0.0637\sym{**} \\
                         &                     &                     &    (0.0142)         &    (0.0309)         \\
\addlinespace
Size                     &     -0.0078         &     -0.0116         &     -0.0088         &     -0.0142         \\
                         &    (0.0084)         &    (0.0190)         &    (0.0084)         &    (0.0189)         \\
\addlinespace
Lev                      &      0.0167         &      0.0126         &      0.0183         &      0.0167         \\
                         &    (0.0316)         &    (0.0722)         &    (0.0315)         &    (0.0720)         \\
\addlinespace
Cashflow                 &      0.0435         &      0.1255\sym{*}  &      0.0415         &      0.1200\sym{*}  \\
                         &    (0.0284)         &    (0.0667)         &    (0.0282)         &    (0.0663)         \\
\addlinespace
Fixed                    &      0.0537         &      0.1405\sym{*}  &      0.0496         &      0.1294         \\
                         &    (0.0354)         &    (0.0812)         &    (0.0354)         &    (0.0812)         \\
\addlinespace
Top1                     &      0.0365         &      0.0754         &      0.0385         &      0.0811         \\
                         &    (0.0519)         &    (0.1156)         &    (0.0515)         &    (0.1147)         \\
\addlinespace
Board                    &      0.0007         &      0.0036         &     -0.0003         &      0.0008         \\
                         &    (0.0203)         &    (0.0486)         &    (0.0203)         &    (0.0486)         \\
\addlinespace
TobinQ                   &      0.0011         &      0.0012         &      0.0009         &      0.0008         \\
                         &    (0.0024)         &    (0.0057)         &    (0.0024)         &    (0.0057)         \\
\addlinespace
\_cons                   &      0.5946\sym{***}&      1.1335\sym{**} &      0.6173\sym{***}&      1.1950\sym{***}\\
                         &    (0.1965)         &    (0.4464)         &    (0.1964)         &    (0.4460)         \\
\midrule
Individual FE            & YES & YES & YES & YES \\
Year FE                  & YES & YES & YES & YES \\
Observations             & 10,849 & 10,849 & 10,849 & 10,849 \\
Adjusted $R^2$           & 0.703  & 0.659  & 0.703  & 0.659  \\
\bottomrule
\end{tabular}
\begin{tablenotes}[para]
  \footnotesize
  \item \textbf{Notes:} This table reports heterogeneity regressions. Columns (1)–(2) include Loss$\times$Certification (Certification=1 if the CSR report is third-party assured); columns (3)–(4) include Loss$\times$WorkSafety(WorkSafety=1 if work-safety content is disclosed). \sym{*}, \sym{**}, \sym{***} denote 10\%, 5\%, 1\% significance, respectively.The standard errors clustered at the firm level are presented in parentheses.
\end{tablenotes}

\end{threeparttable}
\end{adjustbox}
\end{table}

\begin{table}[H]\centering
\def\sym#1{\ifmmode^{#1}\else\(^{#1}\)\fi}
\caption{Heterogeneity of Scale and HighCompetition}
\label{Heterogeneity2}
\begin{adjustbox}{width=\textwidth,center}
\begin{threeparttable}
\begin{tabular}{l*{4}{c}}
\toprule
                         &\multicolumn{1}{c}{(1)}&\multicolumn{1}{c}{(2)}&\multicolumn{1}{c}{(3)}&\multicolumn{1}{c}{(4)}\\
                         &\multicolumn{1}{c}{CSR\_Div\_gini}&\multicolumn{1}{c}{CSR\_Div\_ent}&\multicolumn{1}{c}{CSR\_Div\_gini}&\multicolumn{1}{c}{CSR\_Div\_ent}\\
\midrule
\addlinespace
Loss                   &     -0.0388\sym{***}&     -0.0966\sym{***}&     -0.0003         &     -0.0055         \\
                         &    (0.0099)         &    (0.0230)         &    (0.0082)         &    (0.0188)         \\
\addlinespace
LargeScale                  &     -0.0210\sym{*}  &     -0.0475\sym{*}  &                     &                     \\
                         &    (0.0117)         &    (0.0281)         &                     &                     \\
\addlinespace
Loss$\times$LargeScale           &      0.0297\sym{**} &      0.0732\sym{***}&                     &                     \\
                         &    (0.0117)         &    (0.0278)         &                     &                     \\
\addlinespace
HighCompetition            &                     &                     &      0.0026         &      0.0269         \\
                         &                     &                     &    (0.0207)         &    (0.0495)         \\

\addlinespace
Loss$\times$HighCompetition     &                     &                     &     -0.0289\sym{**} &     -0.0633\sym{**} \\
                         &                     &                     &    (0.0115)         &    (0.0276)         \\
\addlinespace
Size                     &     -0.0046         &     -0.0033         &     -0.0088         &     -0.0143         \\
                         &    (0.0100)         &    (0.0228)         &    (0.0084)         &    (0.0189)         \\
\addlinespace
Lev                      &      0.0288         &      0.0471         &      0.0179         &      0.0166         \\
                         &    (0.0336)         &    (0.0781)         &    (0.0315)         &    (0.0722)         \\
\addlinespace
Cashflow                 &      0.0955\sym{***}&      0.2521\sym{***}&      0.0418         &      0.1196\sym{*}  \\
                         &    (0.0312)         &    (0.0749)         &    (0.0283)         &    (0.0666)         \\
\addlinespace
Fixed                    &      0.0552         &      0.1363         &      0.0508         &      0.1296         \\
                         &    (0.0364)         &    (0.0832)         &    (0.0357)         &    (0.0819)         \\
\addlinespace
Top1                     &      0.0294         &      0.0557         &      0.0358         &      0.0761         \\
                         &    (0.0518)         &    (0.1170)         &    (0.0514)         &    (0.1144)         \\
\addlinespace
Board                    &     -0.0020         &     -0.0027         &     -0.0006         &      0.0006         \\
                         &    (0.0214)         &    (0.0522)         &    (0.0202)         &    (0.0484)         \\
\addlinespace
TobinQ                   &      0.0011         &      0.0010         &      0.0010         &      0.0010         \\
                         &    (0.0028)         &    (0.0064)         &    (0.0024)         &    (0.0057)         \\

\addlinespace
\_cons                   &      0.5320\sym{**} &      0.9691\sym{*}  &      0.6186\sym{***}&      1.1830\sym{***}\\
                         &    (0.2304)         &    (0.5271)         &    (0.1961)         &    (0.4466)         \\
\midrule
Individual FE            & YES & YES & YES & YES \\
Year FE                  & YES & YES & YES & YES \\
Observations             &        9,875         &        9,875         &       10,849         &       10,849         \\
Adjusted $R^2$             &       0.717         &       0.673         &       0.703         &       0.659         \\
\bottomrule
\end{tabular}
\begin{tablenotes}[para]
  \footnotesize
  \item \textbf{Notes:} This table reports heterogeneity regressions by firm scale and product-market competition. Columns (1)–(2) include \emph{Loss$\times$LargeScale} (Scale = 1 for large-scale enterprises); columns (3)–(4) include \emph{Loss$\times$HighCompetition} (HighCompetition = 1 for highly competitive industries). \sym{*}, \sym{**}, \sym{***} denote 10\%, 5\%, and 1\% significance, respectively. Standard errors are clustered at the firm level and reported in parentheses.
\end{tablenotes}

\end{threeparttable}
\end{adjustbox}
\end{table}

%\section*{Acknowledgement}
%This research was supported by the National Innovation Training Program for College Students (project number to be assigned).
\newpage
\bibliography{reference}

\end{document}